%% file: ms.tex
\documentclass[conference]{IEEEtran}

\usepackage{booktabs} % For formal tables

\usepackage{multirow}
\usepackage{listings}
\usepackage{color}
\usepackage{amsmath,amssymb}
\usepackage{url}
\usepackage{algorithm2e}

\usepackage[nounderscore]{syntax}

\newcommand*{\permcomb}[4][0mu]{{{}^{#3}\mkern#1#2_{#4}}}
\newcommand*{\perm}[1][-3mu]{\permcomb[#1]{P}}
\newcommand*{\comb}[1][-1mu]{\permcomb[#1]{C}}
\usepackage{newfloat}
\DeclareFloatingEnvironment[
  fileext   = logr,% don't use log here!
  listname  = {List of Grammars},
  name      = Grammar,
  placement = htp
]{Grammar}

\usepackage{todonotes}
\usepackage{graphicx}
\DeclareGraphicsExtensions{.pdf,.jpeg,.png}

\usepackage{float}   
\newfloat{lablist}{h}{lol}
\floatname{lablist}{List}

\begin{document}

\lstset{language=C}

\lstdefinestyle{nonumbers}
{numbers=none}

\definecolor{mygreen}{rgb}{0,0.4,0}
\definecolor{mygray}{rgb}{0.5,0.5,0.5}
\definecolor{mymauve}{rgb}{0.58,0,0.82}
\lstset{ %
  backgroundcolor=\color{white},   % choose the background color; you
%must add \usepackage{color} or \usepackage{xcolor}
  basicstyle=\ttfamily\small,        % the size of the fonts that are used
%for the code
  basewidth = {.5em, 0.5em},
  breakatwhitespace=false,         % sets if automatic breaks should
%only happen at whitespace
  breaklines=true,                 % sets automatic line breaking
  captionpos=b,                    % sets the caption-position to bottom
  commentstyle=\color{mygreen},    % comment style
  deletekeywords={...},            % if you want to delete keywords from
%the given language
  escapeinside={\%*}{*)},          % if you want to add LaTeX within
%your code
  extendedchars=true,              % lets you use non-ASCII characters;
%for 8-bits encodings only, does not work with UTF-8
  frame=single,	                   % adds a frame around the code
  keepspaces=true,                 % keeps spaces in text, useful for
%keeping indentation of code (possibly needs columns=flexible)
  keywordstyle=\color{blue},       % keyword style
  language=C,                 % the language of the code
  otherkeywords={*,...},           % if you want to add more keywords to
%the set
  numbers=left,                    % where to put the line-numbers;
%possible values are (none, left, right)
  numbersep=5pt,                   % how far the line-numbers are from
%the code
  numberstyle=\tiny\color{black}, % the style that is used for the
%line-numbers
  rulecolor=\color{black},         % if not set, the frame-color may be
%changed on line-breaks within not-black text (e.g. comments (green
%here))
  showspaces=false,                % show spaces everywhere adding
%particular underscores; it overrides 'showstringspaces'
%  showstringspaces=false,          % underline spaces within strings
%only
  showtabs=false,                  % show tabs within strings adding
%particular underscores
  stepnumber=1,                    % the step between two line-numbers.
%If it's 1, each line will be numbered
  stringstyle=\color{mymauve},     % string literal style
  tabsize=2,	                   % sets default tabsize to 2 spaces
%  title=\lstname                   % show the filename of files included
%with \lstinputlisting; also try caption instead of title
  literate={->}{$\rightarrow$}{2}
           {α}{$\alpha$}{1}
           {δ}{$\delta$}{1}
}

\title{Finding Substitutable Binary Code By Synthesizing Adapters}

\author{\IEEEauthorblockN{Vaibhav Sharma}
\IEEEauthorblockA{Department of Computer Science and\\Engineering\\
University of Minnesota\\
Minneapolis, MN 55455\\
vaibhav@umn.edu}
\and
\IEEEauthorblockN{Kesha Hietala}
\IEEEauthorblockA{Department of Computer Science\\
University of Maryland\\
College Park, MD 20742\\
kesha@cs.umd.edu}
\and
\IEEEauthorblockN{Stephen McCamant}
\IEEEauthorblockA{Department of Computer Science and\\Engineering\\
University of Minnesota\\
Minneapolis, MN 55455\\
mccamant@cs.umn.edu}
}
%\author{
% 1st. author
%\alignauthor
%Ben Trovato\\
%       \email{trovato@corporation.com}
%% 2nd. author
%\alignauthor
%G.K.M. Tobin\\
%       \email{webmaster@marysville-ohio.com}
%% 3rd. author
%\alignauthor Lars Th{\o}rv{\"a}ld\\
%       \email{larst@affiliation.org}
%}

\maketitle
\begin{abstract}
Independently developed codebases typically contain many segments of code that perform same or closely related operations (semantic clones). Finding functionally equivalent segments enables applications like replacing a segment by a more efficient or more secure alternative. Such related segments often have different interfaces, so some glue code (an adapter) is needed to replace one with the other. We present an algorithm that searches for replaceable code segments at the function level by attempting to synthesize an adapter between them from some family of adapters; it terminates if it finds no possible adapter. We implement our technique using (1) concrete adapter enumeration based on Intel's Pin framework (2) binary symbolic execution, and explore the relation between size of adapter search space and total search time. We present examples of applying adapter synthesis for improving security and efficiency of binary functions, deobfuscating binary functions, and switching between binary implementations of RC4. We present two large-scale evaluations, (1) we run adapter synthesis on more than 13,000 function pairs from the Linux C library, (2) using more than 61,000 fragments of binary code extracted from a ARM image built for the iPod Nano 2g device and known functions from the VLC media player, we evaluate our adapter synthesis implementation on more than a million synthesis tasks . Our results confirm that several instances of adaptably equivalent binary functions exist in real-world code, and suggest that adapter synthesis can be applied for reverse engineering and for constructing cleaner, less buggy, more efficient programs.
\end{abstract}

% \keywords{program synthesis, semantic clones, symbolic execution, binary analysis}

\input{introduction}
\input{adapter_synthesis}

\input{implementation}

\input{evaluation}

\input{discussion}

\input{related_work}
\input{conclusion}
\bibliographystyle{abbrv}
\bibliography{references}  

\input{new_appendix}

%\listoftodos[Notes]

\end{document}

%% file: introduction.tex
\section{Introduction}
When required to write an implementation for matrix multiplication, the average programmer will come up with a naive implementation in a matter of minutes.
However, much research effort has been invested into creating more efficient matrix multiplication algorithms~\cite{strassen,coppersmith-winograd,gall}.
On attempting to replace the naive implementation with an implementation of a more efficient matrix multiplication algorithm, the programmer is likely to encounter interface differences, such as taking its arguments in a different order.
In this paper we present a technique that automates the process of finding functions that match the behavior specified by an existing function, while also discovering the necessary wrapper needed to handle interface differences between the original and discovered functions.
%
%Our technique takes advantage of the fact that redundancy is an intrinsic property of modular software~\cite{carzaniga}.
%
Other use cases for our technique include replacing insecure dependencies of off-the-shelf libraries with bug-free variants, deobfuscating binary-level functions by comparing their behavior to known implementations, and locating multiple versions of a function to be run in parallel to provide security through diversity~\cite{BorckBDHHJSS2016}, and reverse engineering a fragment of code to its intended semantic functionality.  

%Binary-level semantic clones are often desirable when the user of a insecure off-the-shelf library wishes to replace the library dependency with a bug-free variant.
%%
%But, even after the user successfully finds such a bug-free variant, replacement of the library dependency can only be done if the user also adapts around interface-level differences between the two libraries.
%%
%Reversing engineering tools are a third use case which can benefit from having access to implementations of common functions.
%%
%An obfuscated binary-level function which can be adapted to a simpler implementation allows us to deobfuscate the function in one fell swoop.
%%
%A fourth application is found when binary-level variants of a function are combined to create variants of a program with implementation diversity. 
%%
%Executing this set of diverse variants with the same inputs and monitoring their execution for divergence can provide security through diversity~\cite{BorckBDHHJSS2016}.
%%
%Redundancy is an intrinsic property of modular software and can be also be applied to automatic recovery from runtime failures~\cite{carzaniga}.
%%
%We present a technique which allows us to find binary code that provides equivalent semantics to a target function.

Our technique works by searching for a wrapper that can be added around one function's interface to make it equivalent to another function.
We consider wrappers that transform function arguments and return values.
Listing~\ref{lst:isalpha} shows implementations in two commonly-used libraries of the \textit{isalpha} predicate, which checks if a character is a letter.
Both implementations follow the description of the \textit{isalpha} function as specified in the ISO C standard, but the glibc implementation signifies the input is a letter by returning 1024, while the musl implementation returns 1 in that case.
\begin{lstlisting}[caption={musl and glibc implementations of the \textit{isalpha} predicate and a wrapper around the glibc implementation that is equivalent to the musl implementation}, 
label={lst:isalpha}, style=nonumbers]
int musl_isalpha(int c) {
  return ((unsigned)c|32)-'a' < 26; 
}
int glibc_isalpha(int c) { 
  return table[c] & 1024; 
}
int adapted_isalpha(int c) {
  return (glibc_isalpha(c) != 0) ? 1 : 0; 
}
\end{lstlisting}
The glibc implementation can be adapted to make it equivalent to the musl implementation by replacing its return value, if non-zero, by 1 as shown by the \textit{adapted\_isalpha} function.
This illustrates the driving idea of our approach: to check whether two functions $f_1$ and $f_2$ are different interfaces to the same functionality, we can search for a wrapper that allows $f_1$ to be replaced by $f_2$.

We refer to the function being wrapped around as the {\em inner} function and the function being emulated as the {\em target} function.
In the example above, the inner function is \textit{glibc\_isalpha} and the target function is \textit{musl\_isalpha}.
We refer to the wrapper code automatically synthesized by our tool as an {\em adapter}.
Our adapter synthesis tool searches in the space of all possible adapters allowed by a specified adapter family for an adapter that makes the behavior of the inner function $f_2$ equivalent to that of the target function $f_1$.
We represent that such an adapter exists by the notation $f_1 \leftarrow f_2$.
Note that this adaptability relationship may not be symmetric: $a \leftarrow b$ does not imply $b \leftarrow a$. 
%
%In the \textit{isalpha} example, the other direction of adapter could be implemented by multiplying the musl implementation\rq s return value with 1024, but in general only one direction of adaptation may be possible.
%
To efficiently search for an adapter, we use counterexample guided inductive synthesis~(CEGIS)~\cite{Solar-LezamaTBSS2006}.
An adapter family is represented as a formula for transforming values controlled by parameters: each setting of these parameters represents a possible adapter.
Each step of CEGIS allows us to conclude that either a counterexample exists for the previously hypothesized adapter, or that an adapter exists that will work for all previously found tests.
We use binary symbolic execution both to generate counterexamples and to find new candidate adapters; the symbolic execution engine internally uses a satisfiability modulo theories~(SMT) solver.
We contrast the performance of binary symbolic execution for adapter search with an alternate approach that uses a randomly-ordered enumeration of all possible adapters.
We always restrict our search to a specified finite family of adapters, and also bound the size of function inputs.

We show that adapter synthesis is useful for a variety of software engineering tasks.
One of our automatically synthesized adapters creates adaptable equivalence between a naive implementation of matrix multiplication and an implementation of Strassen\rq s matrix multiplication algorithm.
We also demonstrate the application of adapter synthesis to deobfuscation by deobfuscating a heavily obfuscated implementation of CRC-32 checksum computation.
We find adaptable equivalence modulo a security bug caused by undefined behavior.
Two other pairs of our automatically synthesized adapters create adaptable equivalence between RC4 setup and encryption functions in mbedTLS~(formerly PolarSSL) and OpenSSL.
We can use binary symbolic execution both to generate counterexamples and to find new candidate adapters.
Our notion of adapter correctness only considers code's behavior, so we can detect substitutability between functions that have no syntactic similarity.
We explore the trade-off between using concrete enumeration and binary symbolic execution for adapter search.
Guided by this experiment, we show that binary symbolic execution-based adapter synthesis can be used for reverse engineering at scale.
We use the Rockbox project~\cite{rockbox} to create an a ARM-based 3rd party firmware image for the iPod Nano 2g device and identify more than 61,000 target code fragments from this image.
We extract reference functions from the VLC media player~\cite{vlc}.
Using these target code fragments and reference functions, our evaluation completes more than 1.17 million synthesis tasks.
Each synthesis task navigates an adapter search space of more than 1.353 x $10^{127}$ adapters, enumerating these concretely would take an infeasible amount of time~($10^{14}$ years).
We find our adapter synthesis implementation finds several instances of reference functions in the firmware image. 
Using the most interesting reference functions from this evaluation, we then compare adapter families to explore different parameter settings for adapter synthesis.
%
%
%We verify the correctness of our RC4 adapters by substituting RC4 cipher functions in OpenSSL at the binary level, when they are called by \textit{nmap}~\cite{nmap}, a popular security scanning tool.
%
To test adapter synthesis within the C library, we evaluate two of our adapter families on more than 13,000 pairs of functions from the C library and present synthesized adapters for some of them.
The rest of this paper is organized as follows.
Section~\ref{sec:adapter_synthesis} presents our algorithm for adapter synthesis and describes our adapter families.
Section~\ref{sec:implementation} describes our implementation, and
Section~\ref{sec:evaluation} presents examples of application of adapter synthesis, large-scale evaluations, and a comparison of two adapter search implementations.
Section~\ref{sec:discussion} discusses limitations and future work,
Section~\ref{sec:related-work} describes related work, and
Section~\ref{sec:conclusion} concludes.

%% file: adapter_synthesis.tex
\section{adapter Synthesis} \label{sec:adapter_synthesis}
%First present the counter-example guided adapter synthesis algorithm.
%Then present each of the adapter grammars in BNF.
%Talk about the simple, arithmetic, simple+len, type conversion and
%character-set translation adapter grammars here with an example of each.
%Also mention which operations can be applied to the return value and
%show atleast one motivating example of this application.
%Simple adapter example - wait/waitpid
%Arithmetic adapter portion can be written by Kesha
%simple+len adapter example - already shown in Overview section
%type conversion adapter + type conversion on return value - abs/labs
%return simple+len adapter - itoa example

%To compare functions for equivalence, we try to synthesize an adapter between them that allows us to replace one function by an adapted version of the other.
%
%To synthesize adapters we rely on counterexample-guided search, a technique initially proposed for automatic abstraction refinement by Clarke et al.~\cite{cegar} and later used for program synthesis by Solar-Lezama et al.\cite{Solar-LezamaTBSS2006}. 
%
%In the following sections we present our adapter synthesis algorithm as well as several adapter grammars that we found to be useful for comparing real-world functions for equivalence.
%
\subsection{An Algorithm for adapter Synthesis}
\begin{figure}[]
\centering
\includegraphics[scale=0.34,trim={0 5mm 9cm 0},clip]{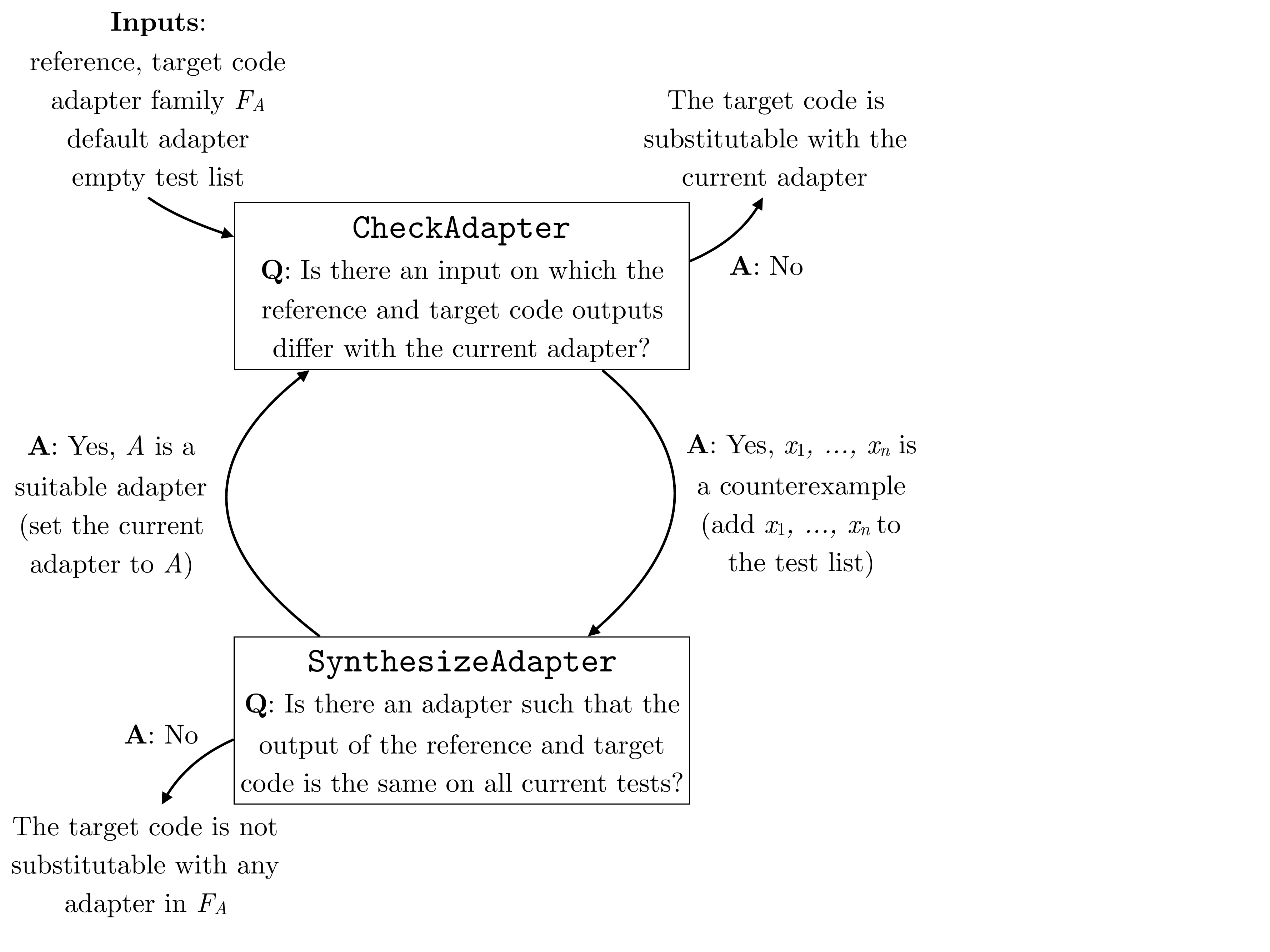}
\caption{Counterexample-guided adapter synthesis}
\label{fig:adapter_synthesis}
\end{figure}

The idea of counterexample-guided synthesis is to alternate between synthesizing candidate adapter expressions, and checking whether those expressions meet the desired specification. 
When a candidate adapter expression fails to meet the specification, a counterexample is produced to guide the next synthesis attempt.
We are interested in synthesizing adapters that map the arguments of the target function to the arguments of the inner function, and map the return value of the inner function to that of the target function, in such a way that the behavior of the two functions match. 
Our specification for synthesis is provided by the behavior of the target function, and we define counterexamples to be inputs on which the behavior of the target and inner functions differ for a given adapter.
Our adapter synthesis algorithm is presented in Algorithm \ref{alg:adapter_search} and is explained in a corresponding figure in Figure~\ref{fig:adapter_synthesis}.

%
%It uses the \textit{CheckAdapter} and \textit{SynthesizeAdapter} procedures which are presented in Algorithms \ref{alg:check_adapter} and \ref{alg:synthesize_adapter}.
%
%Both Algorithms \ref{alg:check_adapter} and \ref{alg:synthesize_adapter} assume side-effects of the target and inner function executions can be captured and checked for equivalence as described in Section \ref{sec:implementation}.
%
%Algorithm \ref{alg:synthesize_adapter} also assumes specification of an adapter grammar that constrains the space of all possible adapters.
%
Algorithm \ref{alg:adapter_search} will terminate with either an adapter that produces equivalence between the target and inner functions for all side-effects, or an indication that the functions cannot be made equivalent using the adapter family we specify.
%
%Algorithm \ref{alg:adapter_search} uses symbolic execution to drive the search for a counter-example during the \textit{CheckAdapter} procedure and to drive the search for an adapter during the \textit{SynthesizeAdapter} procedure. 
%
%Since every execution path explored by symbolic execution must start before or on encountering the first symbolic input, we use symbolic execution to explore execution through target, inner functions in two different ways by marking different inputs of Algorithms \ref{alg:check_adapter}, \ref{alg:synthesize_adapter} symbolic. 
%
\begin{algorithm}[ht]
\LinesNumbered
\small
\SetNlSty{texttt}{[}{]}
\SetKwData{CurrentFadapter}{A}
\SetKwData{CurrentRadapter}{R}
\SetKwData{CEList}{test-list}
\SetKwData{CE}{counterexample}
\SetKwFunction{SynthesizeAdapter}{SynthesizeAdapter}
\SetKwFunction{CheckAdapter}{CheckAdapter}
\SetKwInOut{Input}{Input}\SetKwInOut{Output}{Output}
 \Input{Pointers to the target function T and inner function I}
 \Output{(argument adapter A, return value adapter R) or \textit{null}}
 \CurrentFadapter $\leftarrow$ default-function-args-adapter\;
 \CurrentRadapter $\leftarrow$ default-return-value-adapter\;
 \CEList$\leftarrow$empty-list\;
 \While{true}{
  %input-args $\leftarrow$ list of symbolic variables;\\
  \CE $\leftarrow$ \CheckAdapter(\CurrentFadapter, \CurrentRadapter, T, I);\\
  \eIf{\CE == null}{ 
    \Return{(\CurrentFadapter, \CurrentRadapter)\;} 
  }{ 
    \CEList.append(\CE)\; 
  }
  (\CurrentFadapter, \CurrentRadapter) $\leftarrow$ \SynthesizeAdapter(\CEList, T, I)\;
  \If{\CurrentFadapter == null}{
   \Return{\textit{null}\;}
  }
 }
\caption{Counterexample-guided adapter synthesis}
\label{alg:adapter_search}
\end{algorithm}
\begin{algorithm}[ht]
\small
\LinesNumbered
\SetNlSty{texttt}{[}{]}
\SetKwInOut{Input}{Input}
\SetKwInOut{Output}{Output}
\Input{Concrete adapter A for function arguments and R for return value, target function pointer T, inner function pointer I}
\Output{Counterexample to the given adapter or \textit{null}}
\SetKwData{fadapter}{A}
\SetKwData{radapter}{R}
\SetKwData{args}{args}
\SetKwData{tRet}{T-return-value}
\SetKwData{iRet}{I-return-value}
\SetKwData{tSE}{T-side-effects}
\SetKwData{iSE}{I-side-effects}
\args $\leftarrow$ symbolic\;
\While{execution path available} {
  %execute a path through the target and inner functions using \args\;
  \tRet, \tSE $\leftarrow$ T(\args)\;
  \iRet, \iSE $\leftarrow$ I(\textit{adapt(\fadapter, \args))}\;
  %solve for concrete \args which create behavioral difference between target and inner function using concrete \fadapter\;
  \If{ ! (equivalent(\tSE, \iSE) and equivalent(\tRet, adapt(\radapter, \iRet))) }{
    \Return{concretize(\args)\;}
  }
} 
\Return{null\;}
\caption{CheckAdapter used by Algorithm \ref{alg:adapter_search}}
\label{alg:check_adapter}
\end{algorithm}
%
%
%
%Algorithm \ref{alg:adapter_search} is agnostic to the type and number of side-effects and the adapter grammar.
%
%More expressive adapters allow for equivalence between more function pairs, but may considerably increase the time required to find a correct adapter.
%
%The adapter grammars we present in the next section represent a trade-off between the desires for generality of adapters and reasonable synthesis performance.
%
Algorithm \ref{alg:adapter_search} first initializes the current adapter to a default adapter. 
In our implementation, we often use an `identity adapter' which sets every argument of the inner function to be its corresponding argument to the target function.
Next, the list of current tests is set to be the empty list.
At every iteration (until a correct adapter is found) a new counterexample is added to this list, and any subsequently generated candidate adapter must satisfy all tests in the list.
This provides the intuition for why the adapter search process will always terminate: with every iteration the adapters found become more `correct' in the sense that they produce the desired behavior for more known tests than any previous candidate adapter. 
Algorithm \ref{alg:adapter_search} terminates if the space of candidate adapters allowed by the adapter family is finite.
In practice, we found the number of iterations to be small.
\begin{algorithm}[]
\LinesNumbered
\small
\SetNlSty{texttt}{[}{]}
\SetKwInOut{Input}{Input}
\SetKwInOut{Output}{Output}
\SetKwData{CEList}{test-list}
\SetKwData{CE}{test}
\SetKwData{tRet}{T-return-value}
\SetKwData{iRet}{I-return-value}
\SetKwData{tSE}{T-side-effects}
\SetKwData{iSE}{I-side-effects}
\SetKwData{fadapter}{A}
\SetKwData{radapter}{R}
\SetKwData{eqCounter}{eq-counter}
\Input{List of previously generated counterexamples \CEList, target function pointer T, inner function pointer I}
\Output{(argument adapter \fadapter, return value adapter \radapter) or \textit{null}}
\fadapter $\leftarrow$ symbolic function args adapter\;
\radapter $\leftarrow$ symbolic return value adapter\;
\While{execution path available} {
  \eqCounter $\leftarrow$ 0\;
  \While{\eqCounter $<$ length(\CEList)} {
    %execute a path through the target and inner functions using \CE\;
    \tRet, \tSE $\leftarrow$ T(\CE)\;
    \iRet, \iSE $\leftarrow$ I(adapt(\fadapter, \CE))\;
    %solve for concrete \adapter which gives behavioral equivalence between target and inner function\;
    \eIf{ equivalent(\tSE, \iSE) \textit{and} equivalent(\tRet, adapt(\radapter, \iRet))}{
      \eqCounter $\leftarrow$ \eqCounter + 1\;
    } {break\;}
  }
  \If{ \eqCounter == length(\CEList) }{
    \Return{(concretize(\fadapter), concretize(\radapter))\;}
  }
}
\Return{null\;}
\caption{SynthesizeAdapter used by Algorithm \ref{alg:adapter_search}}
\label{alg:synthesize_adapter}
\end{algorithm}
The \textit{CheckAdapter} procedure (described in Algorithm \ref{alg:check_adapter}) first executes the target function with symbolic arguments and saves its return value and side-effects.
It then plugs in the symbolic function arguments to the concrete adapter given as input to produce adapted symbolic arguments for the inner function by calling the \textit{adapt} method.
Algorithm \ref{alg:check_adapter} executes the adapted inner function, saves its return value and a list of its side-effects.
%
%It does this for every execution path that can be executed first through the target function, and then through the adapted inner function.
%
Algorithm \ref{alg:check_adapter} is successful if it finds inequivalence between (1) side-effects of the target and inner functions, \textbf{or} (2) the target function\rq s return value and adapted return value of the inner function created by calling the \textit{adapt} method with the input return value adapter.
On success, it selects concrete function arguments which create this inequivalence and returns them as a counter-example to the given input adapter.
On failure, it concludes no such inequivalence can be found on any execution path, and returns \textit{null}.

%While Algorithm \ref{alg:check_adapter} finds a counter-example to an adapter, Algorithm \ref{alg:synthesize_adapter} finds an adapter which works for all previously found counter-examples.
%
The \textit{SynthesizeAdapter} procedure described in Algorithm \ref{alg:synthesize_adapter} first concretely executes the target function with a test case from the input test list, and saves the return value side-effects.
It then plugs in the concrete test case into the symbolic argument adapter to create symbolic arguments for the inner function by calling the \textit{adapt} method.
It then executes the inner function, saving its return value and side-effects.
If Algorithm \ref{alg:synthesize_adapter} finds equivalence between (1) side-effects of the target and inner functions, \textbf{and} (2) the target function\rq s return value and the inner function\rq s adapted return value, it considers this test case satisfied.
Finally, on line 15 of Algorithm \ref{alg:synthesize_adapter}, if it finds all tests to be satisifed, it concretizes the function argument and return value adapters and returns them.
The overall time it takes for Algorithm \ref{alg:synthesize_adapter} to find an adapter strongly depends on the space of operations permitted by the adapter family it operates on.
We describe the design of the adapter families, found useful in our evaluation, in the next subsection.
% Too much detail?
%symbolic arguments, saves a set of path constraints introduced by an
%execution path through the target function, and then executes an
%execution path through the adapted inner function which satisfies this set of
%path constraints. 
%This allows the algorithm to prune away execution paths through the
%inner function which are not exercised by inputs to the target function.
%After completing each such execution path extending through the outer
%and adapted inner function, the algorithm calculates if there exists any
%input which produces a mismatch in behavior through side-effects
%observed for the target and inner functions.
%If such an input cannot be found, then the current adapter creates
%equivalence in all observable side-effects of the target and inner
%functions and therefore, is returned as the output of this algorithm on
%line 10.
%If a counter-example can be found, it is added to a list of counter-examples on %line
%7 and moves to line 12.
%The algorithm starts by marking the current adapter as symbolic on line
%12, and then searches for a new adapter that
%creates behavioral equivalence for every currently found counter-example
%on line 13.
%If such an adapter can be found, the algorithm sets it to be the current
%adapter and moves to the counter-example search step on lines 15, 16. 
%If no such adapter was found on line 13, the algorithm terminates with
%an output of inequivalence between the target and inner functions.
%
\subsection{Adapter Families}
\subsubsection{Argument Substitution:}
This family of adapters allows replacement of any inner function argument by one of the target function arguments or a constant.
% Listing \ref{lst:simple_adapter} presents an adapter that can be synthesized using simple argument substitution.
% While this pair of function is not derived from real-world software, this is an interesting example because the functions \textit{$f_1$} and \textit{$f_2$} have a non-intuitive relationship, and it is not immediately obvious that they are equivalent.
This simple family is useful, for instance, when synthesizing adapters between the cluster of functions in the C library that wrap around the \textit{wait} system call as shown in Section \ref{sec:evaluation}.
% 
%\lstinputlisting[caption={Simple argument substitution adapter example}, label={lst:simple_adapter}, style=nonumbers]{code_samples/simple.c}
%
% \textbf{Argument Substitution with String Length:} 
% This family extends the argument substitution adapter family by adding the ability to replace an inner function argument by the length (as computed by \textit{strlen}) of a target function argument. 
% %
% For instance, the C library function \textit{fwrite} can be adapted to \textit{fputs} by setting its second argument to the constant 1 and its third argument to the length of its first argument.\\
% %
\subsubsection{Argument Substitution with Type Conversion:}
This family extends the argument substitution adapter family by allowing inner function arguments to be the result of a type conversion applied to a target function argument. 
Since type information is not available at the binary level, this adapter tries all possible combinations of type conversion on function arguments.
Applying a type conversion at the 64-bit binary level means that each target function argument itself may have been a \textit{char}, \textit{short} or a \textit{int}, thereby using only the low 8, 16, or 32 bytes of the argument register.
The correct corresponding inner function argument could be produced by either a sign extension or zero extension on the low 8, 16, or 32 bits of the argument register. 
%
%Listing \ref{lst:typeconv} presents an additional adapter that can be synthesized for the target and inner functions in Listing \ref{lst:simple_adapter} when type conversions on target function arguments are allowed during adapter search. 
%The $y \& 1$ expression represents a 1-bit zero-extension operation.
This adapter family also allows for converting target function arguments to boolean values by comparing those arguments to zero. 
\subsubsection{Arithmetic adapter:}
This family allows inner function arguments to be arithmetic combinations of target function arguments. 
To ensure that the space of adapters is finite, our implementation only allows for arithmetic expressions of a specified bounded depth.
Arithmetic adapters allow our tool to reproduce other kinds of synthesis.
In the description of the capabilities of the synthesis tool SKETCH, Solar-Lezama et. al.~\cite{Solar-LezamaTBSS2006} present the synthesis of an efficient bit expression that creates a mask isolating the rightmost 0-bit of an integer.
We can synthesize the same bit expression by synthesizing an arithmetic adapter that adapts the identity function to a less-efficient implementation of the operation. 
%
%\lstinputlisting[caption={Type conversion adapter for the function pair shown in Listing \ref{lst:simple_adapter}}, label={lst:typeconv}, style=nonumbers]{code_samples/typeconv.c}
\subsubsection{Memory Substitution:}
\begin{figure}[h]
\caption{Memory substitution adapter to make \textit{struct i} adaptably equivalent to \textit{struct t}}
\label{fig:memsub}
\includegraphics[width=\columnwidth]{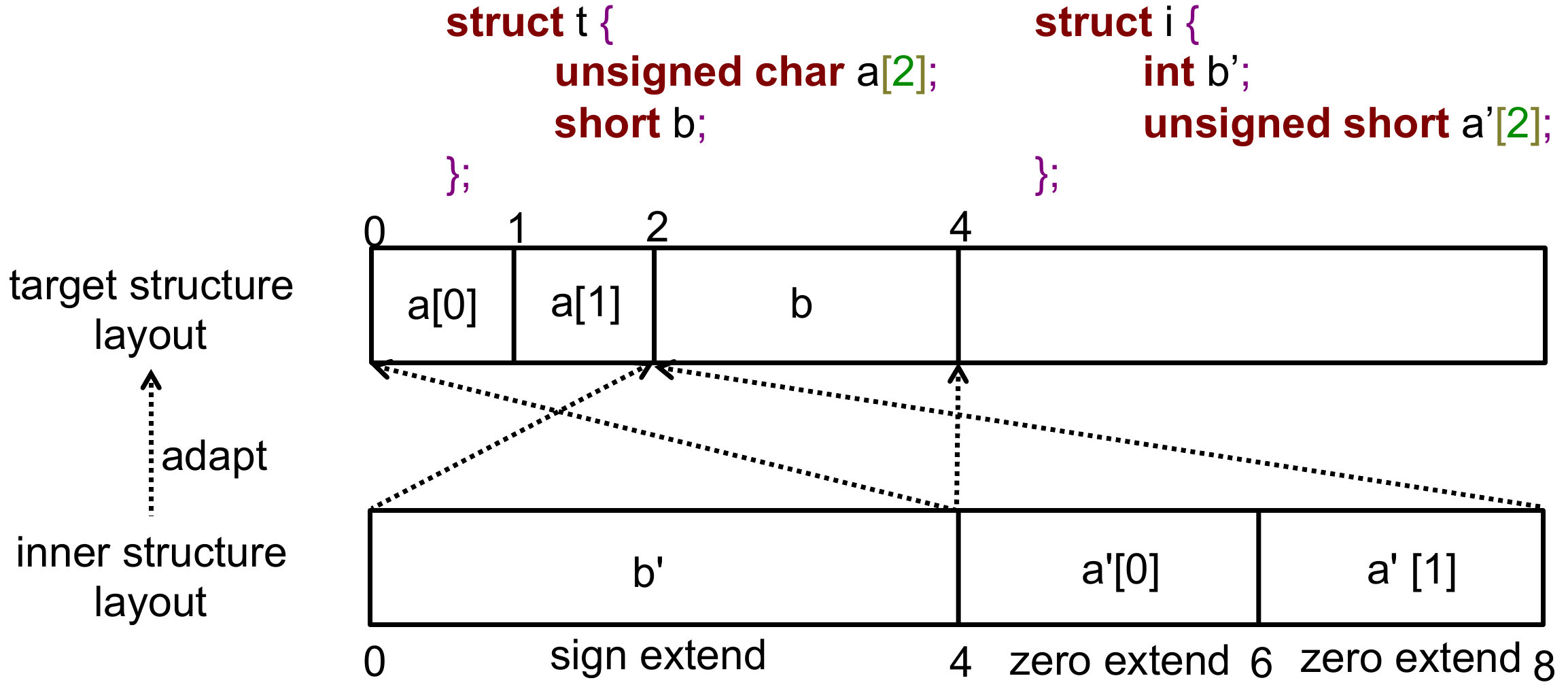}
\end{figure}

This family of adapters allows a field of an inner function structure argument to be adapted to a field of a target function structure argument.
Each field is treated as an array with \textit{n} entries (where n cannot be less than 1), with each entry of size 1, 2, 4, or 8 bytes.
Corresponding array entries used by the target and inner functions need not be at the same address and may also have different sizes, in which case both sign-extension and zero-extension are valid options to explore for synthesizing the correct adapter as shown in Figure~\ref{fig:memsub}.
This makes our adapter synthesis more powerful because it can be used in combination with other rules that allow argument substitution. 
This adapter family synthesizes correct adapters between RC4 implementations in the mbedTLS and OpenSSL libraries in Section~\ref{sec:RC4experiment}. 
\subsubsection{Return Value Substitution:}
The argument substitution families described above can be applied on the return values as well.
An example of different return values having the same semantic meaning is the return value of the C library function \textit{isalpha} as shown in Listing \ref{lst:isalpha}.

\subsection{Example}
\begin{figure}
\begin{lstlisting}[caption={Rockbox iPod Nano implementation of the clamp function followed by a standard C++ implementation of the clamp function.}, label={lst:clamp_source}, style=nonumbers]
int clamp_target(int x) {
    if ((unsigned)x > 255) x = x < 0 ? 0 : 255;
    return x;
}
#include <boost/algorithm/clamp.hpp>
extern "C" int clamp_reference(int x, int lo, int hi)
{
    return boost::algorithm::clamp(x, lo, hi);
}
\end{lstlisting}
\end{figure}
To illustrate our approach, we walk through a representative run of our adapter synthesis algorithm using a target function, that represents binary code from a Rockbox firmware image built for the iPod Nano 2g device, and the clamp function in the Boost library as the reference function. % (shown again in Listing~\ref{lst:clamp_source}).
Both the target code region~(represented as a function) and reference function are shown in Listing~\ref{lst:clamp_source}. 
Although our adapter synthesis implementation can use any binary code region as the target region, in this example we define the target code regions to be a C function, and let inputs correspond to function arguments and output correspond to the function return value. 
Here we will focus only on synthesis of the input adapter, although the general algorithm also produces an adapter that acts the output of the reference function. A correct input adapter should set the first argument of \texttt{clamp\_reference} to the integer argument $x$ of \texttt{clamp\_target} and set the second and third arguments of \texttt{clamp\_reference} to 0 and 255 respectively. We write this adapter as $\mathcal{A}(x) = (x, 0, 255)$.

\textbf{Step 0:} 
adapter synthesis begins with an empty counterexample list and a default adapter that maps every argument to the constant 0 (i.e. $\mathcal{A}(x) = (0,0,0)$). During counterexample generation (\texttt{CheckAdapter} in Figure~\ref{fig:adapter_synthesis}), we use symbolic execution to search for an input $x$ such that the output of \texttt{clamp\_target(x)} is not equivalent to the output of \texttt{clamp\_reference(}$\mathcal{A}(x)$\texttt{)} = \texttt{clamp\_reference(0,0,0)}. From \texttt{CheckAdapter}, we learn that $x = 1$ is one such counterexample.
 
\textbf{Step 1:} Next, during adapter synthesis (\texttt{adapterSynthesis} in Figure~\ref{fig:adapter_synthesis}), we use symbolic execution to search for a new adapter $\mathcal{A}$ that will make \texttt{clamp\_target(x)} equivalent to \texttt{clamp\_reference(}$\mathcal{A}(x)$\texttt{)} for every input $x$ in the list [1]. From \texttt{SynthesizeAdapter}, we learn that $\mathcal{A}(x) = (0,x,x)$ is a suitable adapter, and this becomes our new candidate.

\textbf{Step 2:} At the beginning of this step, the candidate adapter is $\mathcal{A}(x) = (0,x,x)$ and the counterexample list is [1]. First, we use \texttt{CheckAdapter} to search for a counterexample to the current candidate adapter. We find that $x = 509$ is a counterexample. 

\textbf{Step 3:} Next, we use \texttt{SynthesizeAdapter} to search for an adapter $\mathcal{A}$ for which the output of \texttt{clamp\_target(x)} will be equivalent to the output of \texttt{clamp\_reference(}$\mathcal{A}(x)$\texttt{)} for both $x = 1$ and $x = 509$. \texttt{SynthesizeAdapter} identifies $\mathcal{A}(x) = (x,x,255)$ as the new candidate.

\textbf{Step 4:}
At the beginning of this step, the candidate adapter is $\mathcal{A}(x) = (x,x,255)$ and the counterexample list is [1, 509]. As before, first we use \texttt{CheckAdapter} to search for a counterexample to the current candidate adapter. We find that $x = -2147483393$ is a counterexample. 

\textbf{Step 5:} Next, we use \texttt{SynthesizeAdapter} to search for an adapter $\mathcal{A}$ for which the output of \texttt{clamp\_target(x)} will be equivalent to the output of \texttt{clamp\_reference(}$\mathcal{A}(x)$\texttt{)} for every $x \in$ [1, 509, -2147483393]. \texttt{SynthesizeAdapter} identifies $\mathcal{A}(x) = (x, 0, 255)$ as the new candidate. 

\textbf{Step 6:}
In this step, counterexample generation fails to find a counterexample for the current adapter, indicating that the current adapter is correct for all explored paths. Therefore, adapter synthesis terminates with the final adapter $\mathcal{A}(x) = (x, 0, 255)$.
Alternatively, adapter synthesis could have terminated with the decision that the target function is not substitutable by the reference function with any allowed adapter.
In our evaluations, adapter synthesis may also terminate with a timeout, indicating the total runtime has exceeded a predefined threshold.

\subsection{Extensibility}

The adapter synthesis algorithm presented in this section is not tied to any particular source programming language or family of adapters. 
In our implementation (Section~\ref{sec:implementation}) we target binary x86 and ARM code, and we use adapters that allow for common argument structure changes in C code. 
In Section~\ref{sec:evaluation} we present two different interpretations of ``target code regions.''
The first is the function interpretation discussed earlier, where inputs correspond to function arguments and outputs correspond to function return values and side effects.
The second interpretation, enabled by our focus on binary code, defines code regions as ``code fragments.'' 
We define a code fragment to be a sequence of instructions consisting of atleast one instruction. 
Inputs to code fragments are all the general-purpose registers available on the architecture of the code fragment and outputs are registers written to within the code fragment. 
We could also allow reference functions to be more general code regions, but we restricted ourselves to the function-level for now with the idea that a function is the most natural unit of code in which a reverse engineer can express a known behavior.

%% file: implementation.tex
\section{Implementation}\label{sec:implementation}

%\subsection{Tools}
We implement adapter synthesis for Linux/x86-64 binaries using
the symbolic execution tool FuzzBALL~\cite{fuzzball}, which is freely available~\cite{fuzzball-github}.
%Symbolic execution determines what inputs to a program will cause certain behaviors.
%The idea is to execute the program of interest with some concrete values replaced by symbolic variables, and to find satisfying assignments to those symbolic variables that cause the desired execution path to be explored.
FuzzBALL allows us to explore execution paths through the target and adapted inner functions to (1) find counterexamples that invalidate previous candidate adapters and (2) find candidate adapters that create behavioral equivalence for the current set of tests. 
As FuzzBALL symbolically executes a program, it constructs and maintains Vine IR expressions using the BitBlaze~\cite{bitblaze-url} Vine library~\cite{bitblaze-vine} and interfaces with the STP~\cite{stp} decision procedure to solve path conditions.
We replace the symbolic execution-based implementation of adapter search with a concrete implementation that searches the adapter space in a random order.
\subsection{Test Harness}
To compare code for equivalence we use a test harness similar to the one used by Ramos et al.~\cite{Ramos:2011:PLE:2032305.2032360} to compare C functions for direct equivalence using symbolic execution. 
The test harness exercises every execution path that passes first through the target code region, and then through the adapted reference function.
As FuzzBALL executes a path through the target code region, it maintains a path condition that reflects the branches that were taken.
As execution proceeds through the adapted reference function on an execution path, FuzzBALL will only take branches that are do not contradict the path condition.
Thus, symbolic execution through the target and reference code consistently satisfies the same path condition over the input.
Listing \ref{lst:test_harness} provides a representative test harness.
If the target code region is a code fragment, it's inputs $x_1$, ..., $x_n$ need to be written into the first {\tt n} general purpose registers available on the architecture.
Since the target code fragment may write into the stack pointer register ({\tt sp} on ARM), the value of the stack pointer also needs to be saved before executing the target code fragment and restored after the target code fragment has finished execution.
These operations are represented on lines 2, 3, and 5 of Listing \ref{lst:test_harness}.
On line 4 the test harness executes the target code region with inputs $x_1$, ..., $x_n$ and captures its output in {\tt r1}.
If the target code region is a code fragment, its output needs to be determined in a preprocessing phase.
One heuristic for choosing a code fragment\rq s output is to choose the last register that was written into by the code fragment.
On line 9, it calls the adapted reference function \texttt{REF} with inputs $y_1$, ..., $y_m$, which are derived from $x_1$, ..., $x_n$ using an adapter.
It adapts {\tt REF}\rq s return value using the return adapter {\tt R} and saves the adapted return value in {\tt r2}. 
On line 10 the test harness branches on whether the results of the calls to the target and adapted reference code match.
\begin{lstlisting}[caption={Test harness}, label={lst:test_harness}]
void compare(x1, ..., xn) {
  r1 = T(x1, ..., xn); 
  y1 = adapt(A, x1, ..., xn);
  ...
  ym = adapt(A, x1, ..., xn);
  r2 = adapt(R, I(y1, ..., ym));
  if (r1 == r2) printf("Match\n");
  else printf("Mismatch\n");
}

\end{lstlisting}

We use the same test harness for both counterexample search and adapter search. 
During counterexample search, the inputs $x_1$, ..., $x_n$ are marked as symbolic and the adapter is concrete.
FuzzBALL first executes the target code region using the symbolic $x_1$, ..., $x_n$.
It then creates reference function arguments $y_1$, ..., $y_n$ using the concrete adapter and executes the reference function.
During adapter search, the adapter is marked as symbolic, and for each set of test inputs $x_1$, ..., $x_n$, FuzzBALL first executes the target code region concretely.
FuzzBALL then applies symbolic adapter formulas (described in \ref{sec:adapter_formulae}) to the concrete test inputs and passes these symbolic formulas as the adapted reference function arguments $y_1$, ..., $y_n$, before finally executing the reference function.
During counterexample search we are interested in paths that execute the ``Mismatch'' side, and during adapter search we are interested in paths that execute the ``Match'' side of the branches on line 7 of Listing \ref{lst:test_harness}.
For simplicity, Listing \ref{lst:test_harness} shows only the return values $r_1$ and $r_2$ as being used for equivalence checking.
%
%However, during symbolic execution we extend this test harness to do equivalence checking of other state information, including memory and system calls, when comparing the target and adapted reference code executions.
%
\subsection{Adapters as Symbolic Formulae}
\label{sec:adapter_formulae}
\begin{lstlisting}[caption={Argument Substitution adapter}, label={lst:simple_adapter_formula}, style=nonumbers]
y_1_type:reg8_t == 1:reg8_t ? y_1_val:reg64_t : 
  ( y_1_val:reg64_t == 0:reg64_t ? x1:reg64_t : 
    ( y_1_val:reg64_t == 1:reg64_t ? x2:reg64_t : x3:reg64_t ))
\end{lstlisting}
\begin{lstlisting}[label=lst:typeconv_adapter_formula,label={lst:typeconv_adapter_formula},caption={Vine IR formula for one type conversion operation and argument substitution}]
y_1_type:reg8_t == 1:reg8_t ? y_1_val:reg32_t : 
 ( y_1_type:reg8_t == 0:reg32_t ? 
   ( y_1_val:reg32_t == 0:reg32_t ? x1:reg32_t : 
    ( y_1_val:reg32_t == 1:reg32_t ? x2:reg32_t : x3:reg32_t )) 
  : 
   cast( cast( 
     ( y_1_val:reg32_t == 0:reg32_t ? x1:reg32_t : 
      ( y_1_val:reg32_t == 1:reg32_t ? x2:reg32_t : x3:reg32_t )) 
     L:reg16_t ) S:reg32_t )
\end{lstlisting}
We represent adapters in FuzzBALL using Vine IR expressions involving symbolic variables.
As an example, an argument substitution adapter for the adapted inner function argument $y_i$ is represented by a Vine IR expression that indicates whether $y_i$ should be replaced by a constant value (and if so, what constant value) or an argument from the target function (and if so, which argument) 
This symbolic expression uses two symbolic variables, \textit{y\_i\_type} and \textit{y\_i\_val}.
We show an example of an adapter from the argument substitution family represented as a symbolic formula in Vine IR in Listing \ref{lst:simple_adapter_formula}.
This listing assumes the target function takes three arguments, \textit{x1}, \textit{x2}, \textit{x3}.
This adapter substitutes the first adapted inner function argument with either a constant or with one of the three target function arguments.
A value of 1 in \textit{y\_1\_type} indicates the first adapted inner function argument is to be substituted by a constant value given by \textit{y\_1\_val}.
If \textit{y\_1\_type} is set to a value other than 1, the first adapted inner function argument is to be substituted by the target function argument at position present in \textit{y\_1\_val}.
We constrain the range of values \textit{y\_1\_val} can take by adding side conditions. 
In the example shown in Listing \ref{lst:simple_adapter_formula}, when \textit{y\_1\_type} equals a value other than 1, \textit{y\_1\_val} can only equal 0, 1, or 2 since the target function takes 3 arguments.
Symbolic formulae for argument substitution can be extended naturally to more complex adapter families by adding additional symbolic variables.
For example, consider the Vine IR formula shown in Listing~\ref{lst:typeconv_adapter_formula} which extends the formula in Listing~\ref{lst:simple_adapter_formula} to allow sign extension from the low 16 bits of a value.
Listing \ref{lst:typeconv_adapter_formula} begins in the same way as Listing \ref{lst:simple_adapter_formula} on line 1.
But, this time, if \textit{y\_1\_type} is 0, it performs argument substitution based on the value in \textit{y\_1\_val} on lines 3, 4.
If \textit{y\_1\_type} is any value other than 0, it performs sign extension of the low 16 bits in a value.
This value is chosen based on the position set in \textit{y\_1\_val} on lines 8, 9.
Notice lines 8, 9 are the same as lines 3, 4, which means the value, whose low 16 bits are sign-extended, is chosen in exactly the same way as argument substitution.

During the adapter search step of our algorithm, Vine IR representations of adapted inner function arguments are placed into argument registers of the adapted inner function before it begins execution, and placed into the return value register when the inner function returns to the test harness.
When doing adapter synthesis using memory substitution, Vine IR formulas allowing memory substitutions are written into memory pointed to by inner function arguments. 
We use the registers \%rdi, \%rsi, \%rdx, \%rcx, \%r8, and \%r9 for function arguments and the register \%rax for function return value, as specified by the x86-64 ABI calling convention~\cite{x64-abi}.
We do not currently support synthesizing adapters between functions that use arguments passed on the stack, use variable number of arguments, or specify return values in registers other than \%rax.
%
%Our adapter synthesis tool does not support floating point type arguments, but it can be easily extended to allow symbolic formulae on floating point inputs.
%
%While we limit our implementation to synthesize adapters at the function interface level only up to six function arguments, we find a significant portion of the functions in the glibc library are still available for adapter synthesis. 
%
%For synthesizing memory substitution adapters, we write symbolic formulas which allow memory substitution into all addresses used by the target function, up to limited bytes.
%
%Using a smaller limit allows the symbolic formulas to be small and easy to solve but prevents larger structures from being adapted.
%We do not support synthesizing adapters when the function interface of either the target or adapted inner function uses a value of floating point type.
%
\subsection{Equivalence checking of side-effects}
\label{sec:eqchk-syscall}
We record the side-effects of executing the target and adapted inner functions and compare them for equivalence on every execution path.
For equivalence checking of side-effects via system calls, we check the sequence of system calls and their arguments, made by both functions, for equality.
For equivalence checking of side-effects on concretely-addressed memory, we record write operations through both functions and compare the pairs of (address, value) for equivalence.
For equivalence checking of side-effects on memory addressed by symbolic values, we use a FuzzBALL feature called \textit{symbolic regions}. 
Symbolic address expressions encountered during adapted inner function execution are checked for equivalence with those seen during target function execution and mapped to the same symbolic region, if equivalent.
\subsection{Concrete adapter search}
\label{sec:conc-search}
Given an adapter family, the space of possible adapters is finite. 
Instead of using symbolic execution for adapter search, we can concretely check if an adapter produces equal side-effects and return values for all previously-found tests.
We implement concrete enumeration-based adapter search in C for all the adapter families described in Section~\ref{sec:adapter_synthesis}. 
We use the Pin~\cite{pin} framework for checking side-effects on memory and system calls for equality.
To prevent adapter search time from depending on the order of enumeration, we randomize the sequence in which adapters are generated.
%We describe our implementation of equivalence checking on side-effects in more detail in Section \ref{subsec:eqchk-se} in the Appendix. 

%\todo[inline]{Write subsections on equivalence checking for concretely-addressed memory and symbolic regions}

%% file: evaluation.tex
\section{Evaluation}\label{sec:evaluation}
\subsection{Example: Security}
\begin{lstlisting}[caption={two implementations for mapping ordered keys,negative or positive, to values using a C array}, 
label={lst:lookup}]
int l1(int *table, int len, int c) { 
  if(abs(c) > (len/2)) // bug if c = -2147483648
    return -1;  
  return table[c+(len/2)+1];
}
int l2(int c, int *table, int len) {
  if( !(-(len/2) <= c && c <= (len/2)) ) 
    return -1;
  return table[c+(len/2)+1];
}
\end{lstlisting}
%
%Data structures that map keys to their corresponding values are commonly found in modern programming languages~\cite{hashtbl-ocaml},~\cite{map-cpp}. 
%
%Support for such data structures is not part of the C language standard.
%
%Arrays in C provide a convenient and fast way for mapping a previously known number of integer keys to values.
%
%Implementation of data compression algorithms and signal processing implementations require constant-time lookup for negative and positive powers of values.
%
%In such situations, an array in C can be the perfect solution.
%
Consider a table implementing a function of a signed input.
For example, keys ranging from -127 to 127 can be mapped to a 255-element array.
Any key \textit{k} will then be mapped to the element at position \textit{k}+127 in this array.
We present two implementations of such lookup functions in Listing~\ref{lst:lookup}.
Both functions, \textit{l1} and \textit{l2}, assume keys ranging from -\textit{len}/2 to +\textit{len}/2 are mapped in the \textit{table} parameter.
However, \textit{l1} contains a bug caused due to undefined behavior.
The return value of \textit{abs} for the most negative 32-bit integer~(-2147483648)~is not defined~\cite{gnu-abs}.
The eglibc-2.19 implementation of \textit{abs} returns the absolute value of the most negative 32-bit integer as this same 32-bit integer.
This causes the check on line 2 of Listing~\ref{lst:lookup} to not be satisfied allowing execution to continue to line 4 and cause a segmentation fault.
Even worse, passing a carefully-chosen value for \textit{len} can allow a sensitive value to be read and allow this bug to be exploited by an attacker.
\textit{l2} in Listing~\ref{lst:lookup} performs a check, semantically-equivalent to the one on line 2, but does not contain this bug.
Our adapter synthesis implementations were able to synthesize correct argument substitution adapters in the \textit{l1} $\leftarrow$ \textit{l2} direction.
adapter synthesis with concrete enumeration-based adapter search takes 5 seconds, and with FuzzBALL-based adapter search takes 41 seconds.
This adapter synthesis requires adaptation modulo the potential segmentation fault in \textit{l1}. 
This example shows adapter synthesis provides replacement of buggy functions with their bug-free siblings by adapting the interface of the bug-free function to the buggy one.
\subsection{Example: Deobfuscation}
A new family of banking malware named Shifu was reported in 2015~\cite{fireeye-shifu},~\cite{ibm-shifu}.
Shifu was found to be targeting banking entities across the UK and Japan. 
It continues to be updated~\cite{paloalto-shifu}.
Shifu is heavily obfuscated, and one computation used frequently in Shifu is the computation of CRC-32 checksums.
We did not have access to the real malware binary, but we were able to simulate its obfuscated checksum computation binary function using freely-available obfuscation tools.

\begin{figure}[]
\caption{Argument substitution adapter to make one obfuscated CRC-32 checksum function adaptably equivalent to the reference function}
\label{fig:obfs}
\includegraphics[width=\columnwidth]{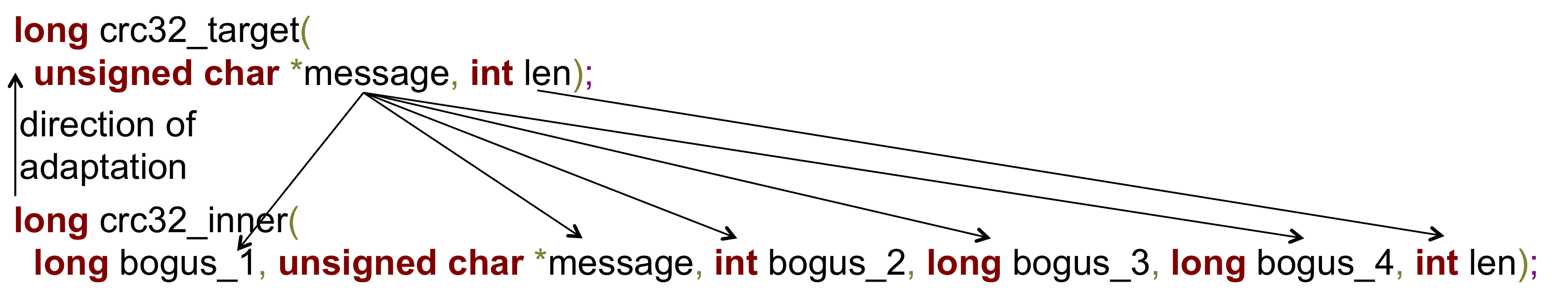}
\end{figure}

Given a reference implementation of CRC-32 checksum computation, adapter synthesis can be used to check if an obfuscated implementation is adaptably equivalent to the reference implementation.
We used the implementation of CRC-32 checksum computation available on the adjunct website~\cite{hd-crc} of Hacker\rq s Delight~\cite{hd-book} (modified so that we could provide the length of the input string) as our reference function.
We obfuscated this function at the source code and Intermediate Representation~(IR) levels to create three obfuscated clones.
For the first clone, we used a tool named Tigress~\cite{tigress} to apply the following source-level obfuscating transformations: 
\begin{enumerate}
\item Function virtualization: This transformation turns the reference function into an interpreter with its own bytecode language.
\item Just-in-time compilation/dynamic unpacking: This transformation translates each function \textit{f} into function \textit{f'} consisting of intermediate code so that, when \textit{f'} is executed, \textit{f} is dynamically compiled to machine code.
\item reordering the function arguments randomly, inserting bogus arguments, adding bogus non-trivial functions and loops, and allowing superoperators~\cite{superoperators}. 
\end{enumerate}
These transformations led to a 250\% increase in the number of source code lines.
For the second clone, we applied the following obfuscating transformations at the LLVM IR level using Obfuscator-LLVM~\cite{obfs-llvm}:
\begin{enumerate}
\item Instruction Substitution: This transformation replaces standard binary operators like addition, subtraction, and boolean operators with functionally equivalent, but more complicated, instruction sequences.
\item Bogus Control Flow: This transformation modifies the function call graph by injecting basic blocks for bogus control flow and modifying existing basic blocks by adding junk instructions chosen at random. 
\item Control flow flattening: This transformation flattens the control flow graph of the clone in a way similar to L{\'a}szl{\'o} et al~\cite{fla}.
\end{enumerate}
These transformations caused the number of instruction bytes to increase from 126 to 2944 bytes.
Finally, we compiled the obfuscated C code (obtained using Tigress) with the LLVM obfuscator tool to create a third clone.
We then ran our adapter synthesis tool with the reference function as the target function and all three clones as inner functions.
We used the CRC-32 checksum of a symbolic 1 byte input string as the return value of each clone.
Our adapter synthesis tool, using FuzzBALL-based adapter search, correctly concluded that all three clones were adaptably equivalent to the reference function in less than 3 minutes using argument substitution.
A correct adapter for one obfuscated clone is shown in Figure~\ref{fig:obfs}. 
It maps the string and length arguments correctly, while ignoring the four bogus arguments (the mappings to bogus arguments are irrelevant).
While performing adapter synthesis on heavily-obfuscated binary code is challenging, adaptation in this example is complicated further by an increase in the number of inner function arguments causing the adapter search space to increase to 43.68 million adapters. 
%
%
% We confirmed this by hand-crafting a new adapter that assigned a constant value of 0 to the bogus arguments, leaving the mapping for the string and length arguments intact, and ran our counter-example search against this adapter.
% %
% Our counter-example search quickly concluded that this adapter is also correct.
% %
% We also ran the same adapter synthesis experiments multiple times with varying random seeds.
% %
% adapters synthesized in these rounds maintained the mapping for the string and length arguments, and assigned other random constant values to the bogus arguments.
%
\subsection{Example: Efficiency}
\begin{lstlisting}[caption={Naive implementation of matrix multiplication}, 
label={lst:naive_mm}, style=nonumbers]
#define N 4
void naive_mm(int** C, int** A, int** B) {
  int i, j, k;
  for ( i = 0; i < N; i++)
    for ( j = 0; j < N; j++) {
      C[i][j] = 0;
      for ( k = 0; k < N; k++)
        C[i][j] += A[i][k]*B[k][j];
    }
}
\end{lstlisting}
Adapter synthesis can also be applied to find more efficient versions of a function, even when those versions have a different interface.
Matrix multiplication is one of the most frequently-used operations in mathematics and computer science. 
It can be used for other crucial matrix operations~(for example, gaussian elimination, LU decomposition~\cite{algorithms}) and as a subroutine in other fast algorithms~(for example, for graph transitive closure).
Adapting faster binary implementations of matrix multiplication to the naive one\rq s interface improves the runtime of such other operations relying on matrix multiplication.
Hence, as our target function, we use the naive implementation of matrix multiplication shown in Listing~\ref{lst:naive_mm}.
As our inner function we use an implementation of Strassen\rq s algorithm~\cite{strassen} from Intel\rq s website~\cite{intel-strassen}, which takes the input matrices \textit{A} and \textit{B} as the 1st and 2nd arguments respectively and places the product matrix in its 3rd argument.
We modified their implementation so that it used Strassen's algorithm for all matrix sizes.
Our adapter synthesis tool, using FuzzBALL-based adapter search, finds the correct argument substitution adapter for making the implementation using Strassen's algorithm adaptably equivalent to the naive implementation in 17.7 minutes for matrices of size 4x4.
When using concrete enumeration-based adapter search, the adapter search finds the correct adapter in less than 4.5 minutes.
%
%Of these 17.7 minutes, the final counterexample search step takes about 8 minutes because it is searching for a counterexample to the correct adapter.
%
%We verified the correctness of our adapter by using it with random matrices of sizes upto 1024x1024 and checking the resulting product matrix for equality with that produced by the naive implementation.
%

This example shows that adapter synthesis can be used for finding adaptably equivalent clones of a function that have different algorithmic space and time complexity.
Program analysis techniques for checking space and time usage of different implementations are being actively researched~\cite{darpa-stac}.
Symbolic execution can also be used for finding inputs that cause functions to exhibit worst-case computational complexity~\cite{wise}.
adapter synthesis can be used as a pre-processing step before applying other techniques for detecting algorithmic complexity of semantic clones.

\subsection{Example: RC4 encryption} \label{sec:RC4experiment}
To show that adapter synthesis can be applied to replace one library with another, we chose to adapt functions implementing RC4 functionality in mbedTLS and OpenSSL.
\noindent
\subsubsection{RC4 key structure initialization:} 
\begin{figure}[]
\caption{Argument substitution adapter to make \textit{RC4\_set\_key} adaptably equivalent to \textit{mbedtls\_arc4\_setup}}
\label{fig:rc4setup_adapter}
\includegraphics[width=\columnwidth]{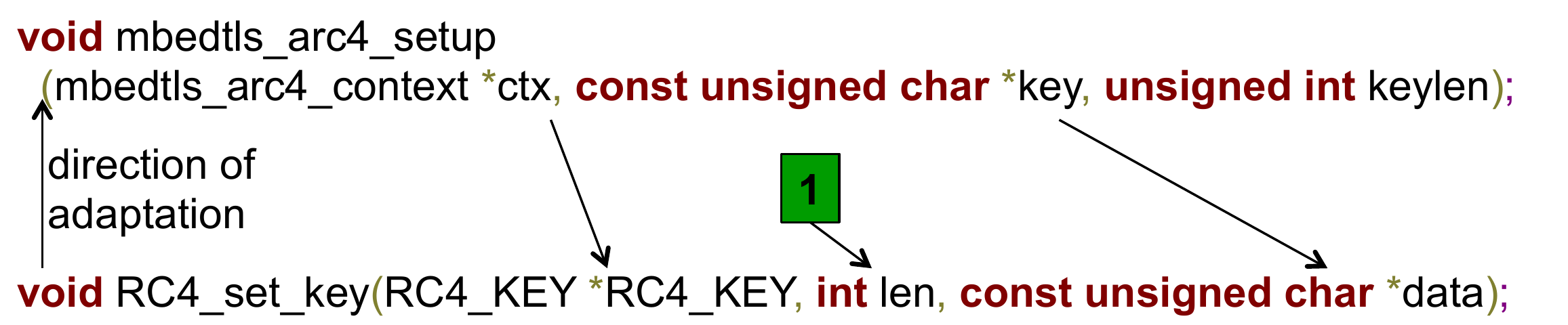}
\end{figure}

The RC4 algorithm uses a variable length input key to initialize a table with 256 entries within the key structure argument.
Both cryptography libraries in our example, mbedTLS and OpenSSL, have their own implementation of this initialization routine. Both function signatures are shown in Figure \ref{fig:rc4setup_adapter}. 
Executing each of these initialization routines requires 256 rounds of mixing bytes from the key string into the key structure.
%
%Each round requires two load and two store operations into an array with 256 entries.
%
The two initialization routines require the key length argument at different positions, so making \textit{RC4\_set\_key} adaptably equivalent to \textit{mbedtls\_arc4\_setup} requires not only mapping the \textit{mbedtls\_arc4\_context} object to a \textit{RC4\_KEY} object, but also figuring out the correct mapping of the key length argument.
This combination of argument substitution and memory substitution adapter families creates a search space of 421.823 million adapters. 

Our adapter synthesis tool correctly figures out both mappings and finds adaptable equivalence by creating equivalence between side-effects on the structure objects~(\textit{ctx} for \textit{mbedtls\_arc4\_setup}, \textit{RC4\_KEY} for \textit{RC4\_set\_key}).
To setup adapter synthesis between these two function pairs (we synthesized adapters in both directions), we used a symbolic key string of length 1, and hence the synthesis tool correctly sets the key length argument to 1.
Our tool, when using FuzzBALL-based adapter search, figures out the correct memory and argument substitution adapters in the mbedTLS $\leftarrow$ OpenSSL direction for initialization routines in 60 minutes and in the OpenSSL $\leftarrow$ mbedTLS direction in 49 minutes.
Thus, we combined the memory substitution adapter with the argument substitution adapter family to synthesize adaptable equivalence between the RC4 setup pair of functions.
% Our adapter synthesis tool finds the correct adapter for RC4 encryption in its second adapter search step, but finds the right adapter for RC4 initialization in the first adapter search step itself.
% %
% This second adapter search step, in case of RC4 encryption, took 68 minutes, which is why our tool finished faster in case of RC4 initialization.
% %
% This observation exhibits the phenomenon that adapter search gets more expensive as the inner function is to be adapted for more test inputs.
% %
% But, adaptation for more test inputs also causes the adapter search to be moved in the right direction.
\noindent
\subsubsection{RC4 encryption:} 
\begin{figure}[]
\caption{Memory substitution adapter to make \textit{RC4\_KEY} adaptably equivalent to \textit{mbedtls\_arc4\_context}}
\label{fig:rc4_struct_adapter}
\includegraphics[width=\columnwidth]{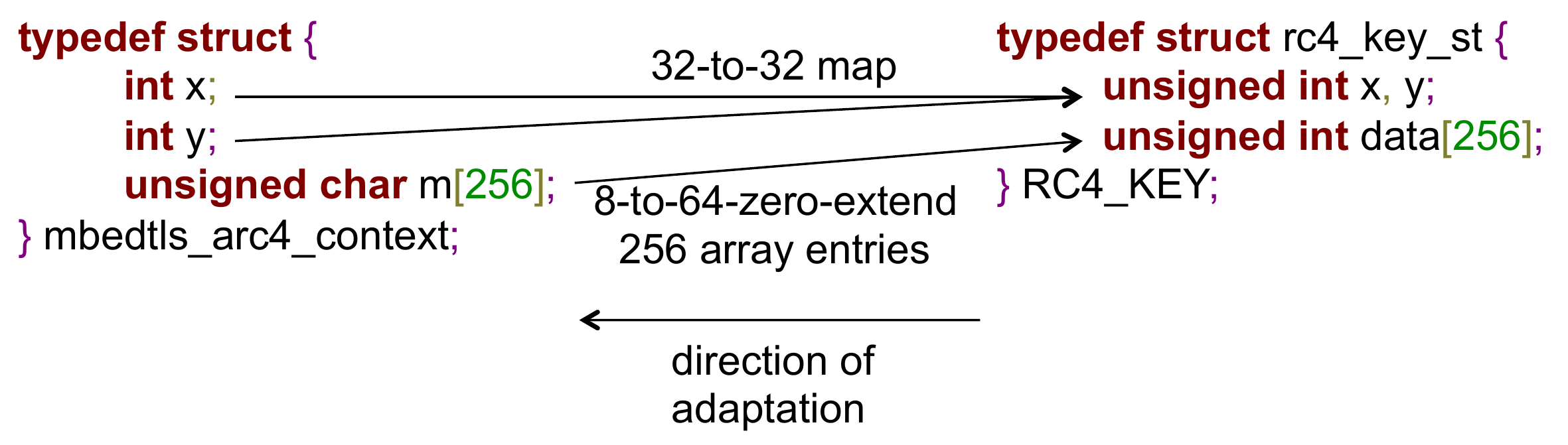}
\end{figure}

RC4 encryption functions in mbedTLS and OpenSSL take 4 arguments each, one of which is the RC4 key structure argument.
The RC4 key structures~(\textit{RC4} in OpenSSL, \textit{mbedtls\_arc4\_context} in mbedTLS) contain three fields as shown in Figure \ref{fig:rc4_struct_adapter}. 
The first two 4-byte fields are used to index into the third field, which is an array with 256 entries.
Each entry is 4 bytes long in OpenSSL and 1 byte long in mbedTLS. 
In order to present an example of a memory substitution adapter synthesized in isolation, we created wrappers for both RC4 encryption functions so that only the key structure argument was exposed and used a fixed value for the input string.
%
%We captured the output of the encryption using side-effects on memory.
%
This allowed us to direct the adapter search to search for all possible mappings between the mbedTLS and OpenSSL RC4 key structure fields.
Allowing arbitrary numbers of 1, 2, 4, or 8 byte entries in each field of the 264~(2*4+256*1) byte mbedTLS key structure and 1032~(2*4+256*4) byte OpenSSL key structure made the search space of memory mappings very large, so we instead only explored adapters where the number of entries in each array was a power of 2.
While this reduction is useful in practice, it still gives us a search space of about 4.7 million adapters in both directions of adaptation.

The correct adapter that adapts the OpenSSL key structure to the mbedTLS key structure~(mbedTLS $\leftarrow$ OpenSSL) performs 2 mapping operations: (1) it maps the first 2 mbedTLS key structure fields directly to the first 2 OpenSSL key structure fields and (2) it zero extends each 1 byte entry in the 3rd field of the mbedTLS key structure to the corresponding 4 byte entry in the 3rd field of the OpenSSL key structure.
The correct adapter for adapting in the reverse direction~(OpenSSL $\leftarrow$ mbedTLS) changes the second mapping operation to map the least significant byte of each 4 byte entry in the 3rd field to the 1 byte entry in its corresponding position.
Our adapter synthesis tool, when using FuzzBALL-based adapter search, found the correct memory substitution adapter in the mbedTLS $\leftarrow$ OpenSSL direction in 2.4 hours and in the OpenSSL $\leftarrow$ mbedTLS direction in 2.6 hours.
When using concrete enumeration-based adapter search, we found the correct adapter in 1.8 hours in the mbedTLS $\leftarrow$ OpenSSL direction, of which only 6 minutes were spent on adapter search.
In the OpenSSL $\leftarrow$ mbedTLS direction, we found the correct adapter, with concrete enumeration-based adapter search, in 65 minutes, of which only 1.5 minutes were spent on adapter search.
The correct adapter for making \textit{RC4\_KEY} adaptably equivalent to \textit{mbedtls\_arc4\_context} is shown in Figure \ref{fig:rc4_struct_adapter}.
%
%During our evaluation, our tool found the correct adapter in the OpenSSL $\leftarrow$ mbedTLS and mbedTLS $\leftarrow$ OpenSSL directions in 2.7 hours using a input string of length 1.
%
We verified the correctness of our adapted key structures by using self-tests present in mbedTLS and OpenSSL. 
%
%We also verified our memory substitution adapters to be correct for symbolic input strings of length 2.
%
%Given the design of the RC4 encryption algorithm, we do not expect the correctness of our adapters to change for longer input strings.
\noindent
\subsubsection{RC4 adapter verification using nmap:} 
\begin{figure}[]
\caption{nmap using RC4 encryption in mbedTLS instead of OpenSSL}
\label{fig:nmap_struct_adapter}
\includegraphics[width=\columnwidth]{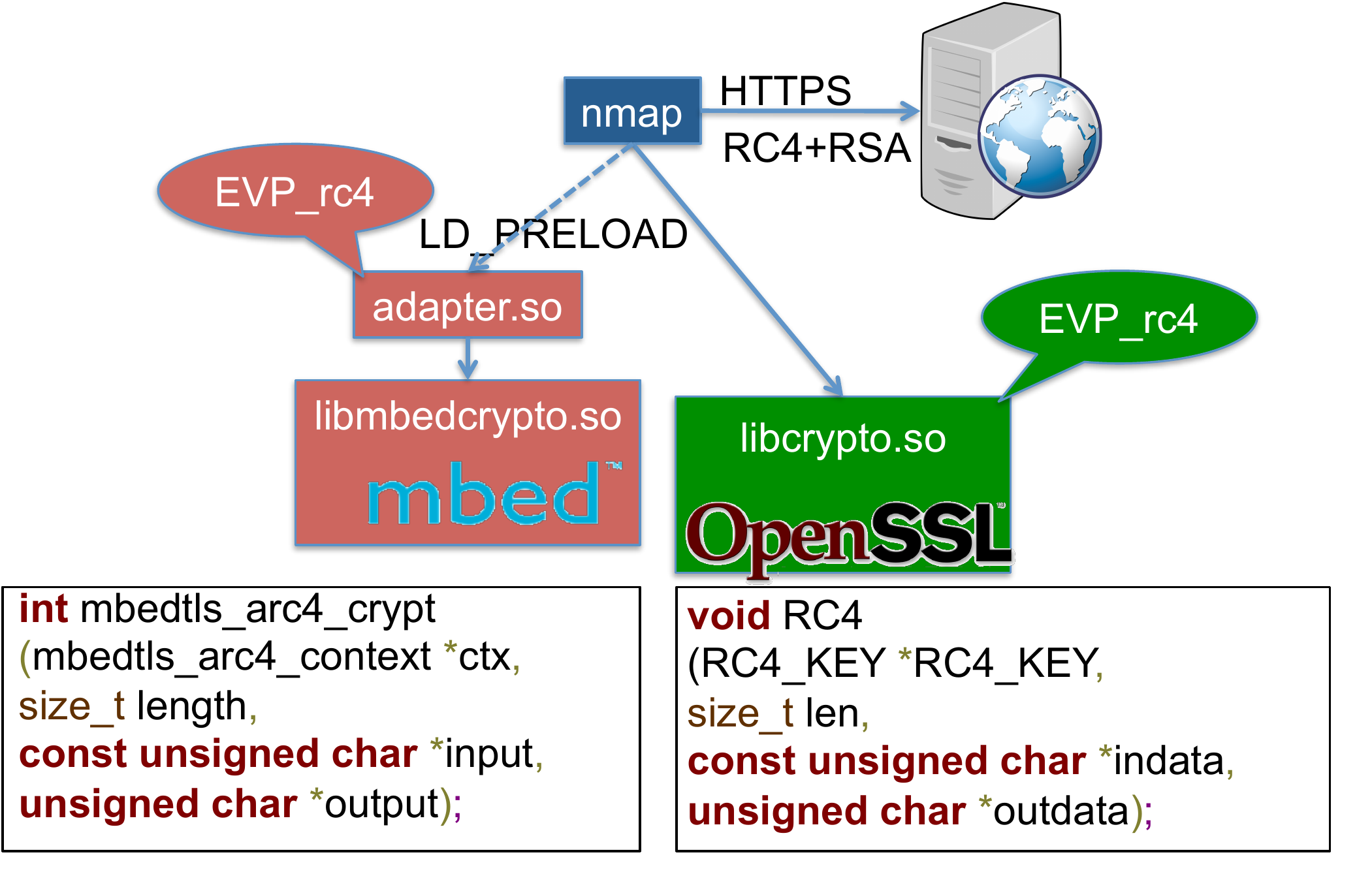}
\end{figure}

We verified our RC4 memory substitution adapter using nmap, as shown in Figure \ref{fig:nmap_struct_adapter}.
We created adapted versions of the OpenSSL RC4 setup and encryption functions that internally use the mbedTLS key structure adapted to the OpenSSL key structure.
On a 64-bit virtual machine running Ubuntu 14.04, we compiled the adapted setup and encryption functions into a shared library and setup a local webserver on the virtual machine, which communicated over port 443 using the \textit{RC4+RSA} cipher.
We used the stock nmap binary to scan our localhost and injected our specially created shared library using the \textit{LD\_PRELOAD} environment variable.
The preloading caused the RC4 functions in our shared library to be executed instead of the ones inside OpenSSL. 
The output of nmap, run with preloading our specially created shared library which used the OpenSSL $\leftarrow$ mbedTLS structure adapter, was the same as the output of nmap which used the system OpenSSL library.
\subsection{Evaluation with C library}
\subsubsection{Setup}
We evaluated our adapter synthesis tool on the system C library available on Ubuntu 14.04~(eglibc 2.19).
The C library uses a significant amount of inlined assembly, for instance, the \textit{ffs}, \textit{ffsl}, \textit{ffsll} functions, which
motivates automated adapter synthesis at the binary level.
We enumerated 1316 exported functions in the library in the order they
appear, which caused functions that are defined in the same source files
to appear close to each other.
Considering every function in this list as the target function, we chose five functions that appear above and below it as 10 potential inner functions.
These steps gave us a list of 13130~(10$\times$1316 - 2$\times$ $\sum_{i=1}^5 i$) pairs of target and inner functions.
%since the first five and the last five target functions 
%had less than five potential semantically-equivalent inner functions. 
%
We used the argument substitution and type conversion adapter families combined with the return value adapter family because these families scale well and are widely applicable.
We ran our adapter synthesis with a 2 minute timeout on a machine running CentOS 6.8 with 64-bit Linux kernel version 2.6.32 using 64 GB RAM and a Intel Xeon E5-2680v3 processor.
To keep the running time of the entire adapter synthesis process within practical limits, we configured FuzzBALL to use a 5 second SMT solver timeout and to consider any queries that trigger this timeout as unsatisfiable.
We limited the maximum number of times any instruction can be executed to 4000 because this allowed execution of code which loaded library dependencies.
We limited regions to be symbolic up to a 936 byte offset limit (the size of the largest structure in the library interface) and any offset outside this range was considered to contain zero.
\subsubsection{Results}
Table~\ref{table:libc-evaluation} summarizes the results of searching for argument substitution and type conversion adapters with a return value adapter within the 13130 function pairs described above.
The similarity in the results for the type conversion adapter family and argument substitution adapter family arises from the similarity of these two families.
%
%We synthesized the argument substitition adapter, with and without type conversion, along with the return value adapter.
%However, trying to synthesize an adapter using a simpler grammar translates into simpler queries for the solver, which in turn, results in the adapter synthesis tool concluding with a result of equivalence or the lack of it within the two minute hard timeout.
%
%For about 69\% of the 13130 function pairs, our synthesis tool concluded that functional equivalence cannot be created between the target and inner functions for the chosen adapter grammar.
%
%Our adapter synthesis tool timed out for about 21\% of the function pairs.
%
The most common causes of crashing during execution of the target function were missing system call support in FuzzBALL, and incorrect null dereferences~(caused due to lack of proper initialization of pointer arguments).
The timeout column includes all function pairs for which we had a solver timeout~(5 seconds), hit the iteration limit~(4000), or reached a hard timeout~(2 minutes).
The search terminated without a timeout for 70\% of the function pairs, which reflects a complete exploration of the space of symbolic inputs to a function, or of adapters.
%Because our test harness executes the target function first, in some cases, FuzzBALL failed to execute any execution paths to completion before starting execution of the inner function.
%
%While missing system call functionality in FuzzBALL was one cause for these failures, FuzzBALL often incorrectly classified pointer argument dereferences in the target function as null dereferences which caused the execution path to terminate.
%
%Another cause of failure to execute any execution paths through the target function to completion was the iteration limit of 4000 we used during our experiments.
%
%Our adapter synthesis tool reports 391 and 392 adapters for the argument substition grammar without type conversion and with type conversion respectively.
%
%
%
%We present the true positives found by our C library evaluation in the next subsection.
%
%But, during the last counter-example search step for the reported adapter, not every execution path completed execution through the target function.
%
%If a counter-example search for an adapter finishes executing the target function during every execution path and still fails to find a counter-example, it guarantees that no 
%
%Therefore, we report the number of adapters - 175 out of 391 in case of simple and 174 out of 392 in case of type conversion adapters - for which every execution path during the the final counter-example search step completed execution through the target function.
\input{adapter_results_1}

% One of the the most interesting conclusions from our adapter synthesis tool
% was the adapters reported between the 13130 function pairs, of which
% 2.8\% and 2.7\% were reported to have an adapter when using the argument
% substitution grammar without type conversion and with type conversion respectively.
%
%
Since there is no ground truth, we manually corroborated the results of our evaluation by checking the C library documentation and source code.
Our adapter synthesis evaluation on the C library reported 30 interesting true positives shown in Table \ref{table:libc-adapters}.
(The remaining adapters found are correct but trivial.)
The first column shows the function pair between which an adapter was
found (with the number of arguments) and the second
column shows the adapters.
We use the following notation to describe adapters in a compact way.
$f_1$ $\leftrightarrow$ $f_2$ means $f_1$ $\leftarrow$ $f_2$ and $f_2$ $\leftarrow$ $f_1$.
\# followed by a number indicates inner argument substitution by a target argument, while other numbers indicate constants.
X-to-YS represents taking the low X bits and sign extending them to Y bits, X-to-YZ represents a similar operation using zero extension.
The last three rows shown in Table \ref{table:libc-adapters} shows three arithmetic adapters found within the C library using partial automation.
We synthesized the correct adapters by writing wrappers containing preconditions around the \textit{isupper}, \textit{islower}, \textit{kill}  functions.
\subsection{Comparison with Concrete Enumeration-based Adapter Search}
\begin{figure}[]
\caption{Comparing concrete enumeration-based adapter search with binary symbolic execution-based adapter search}
\label{fig:conc_vs_se}
\includegraphics[width=\columnwidth]{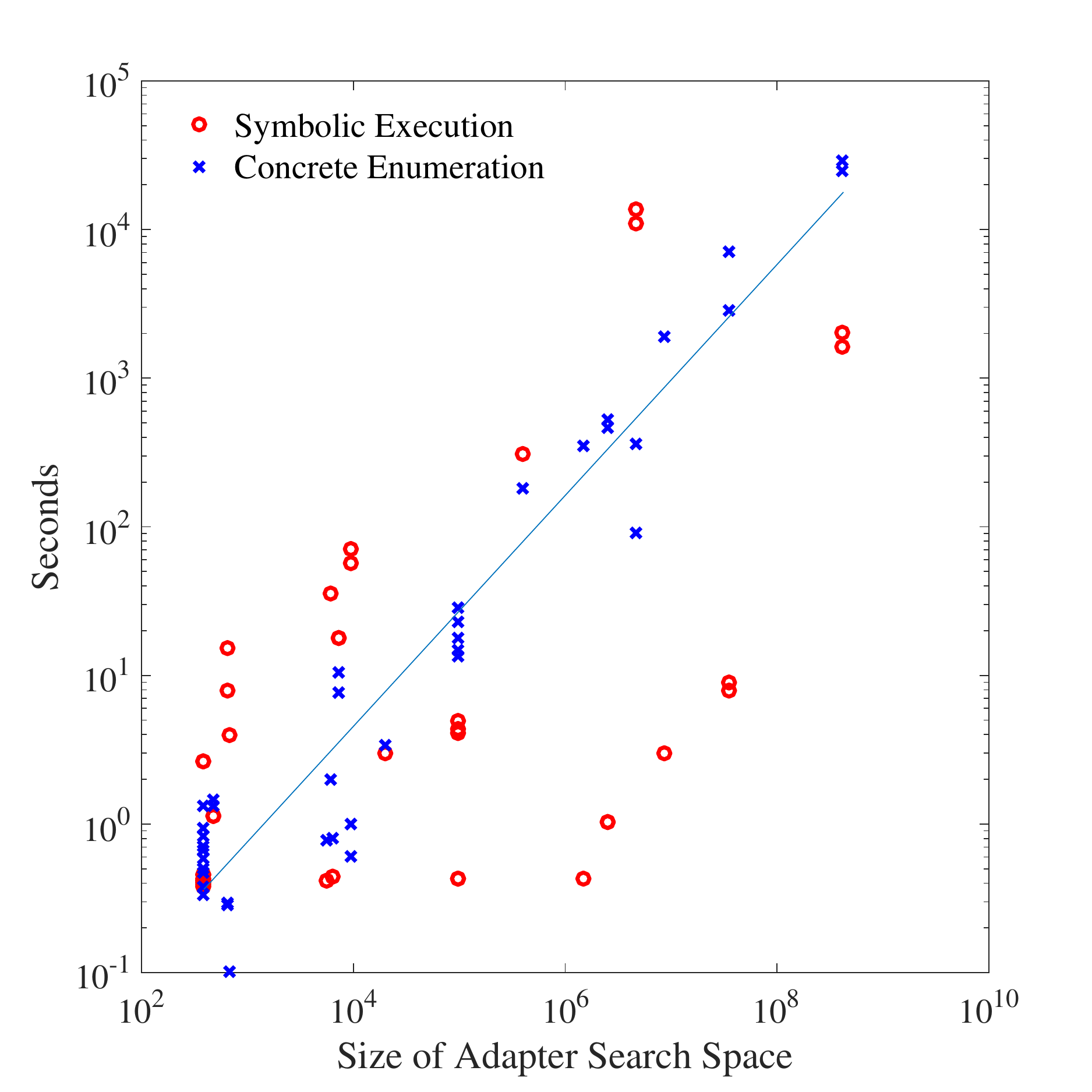}
\end{figure}

\begin{figure}[]
\caption{Comparing concrete enumeration-based adapter search with binary symbolic execution-based adapter search using variants of \textit{popCnt}}
\label{fig:popCnt}
\includegraphics[width=\columnwidth]{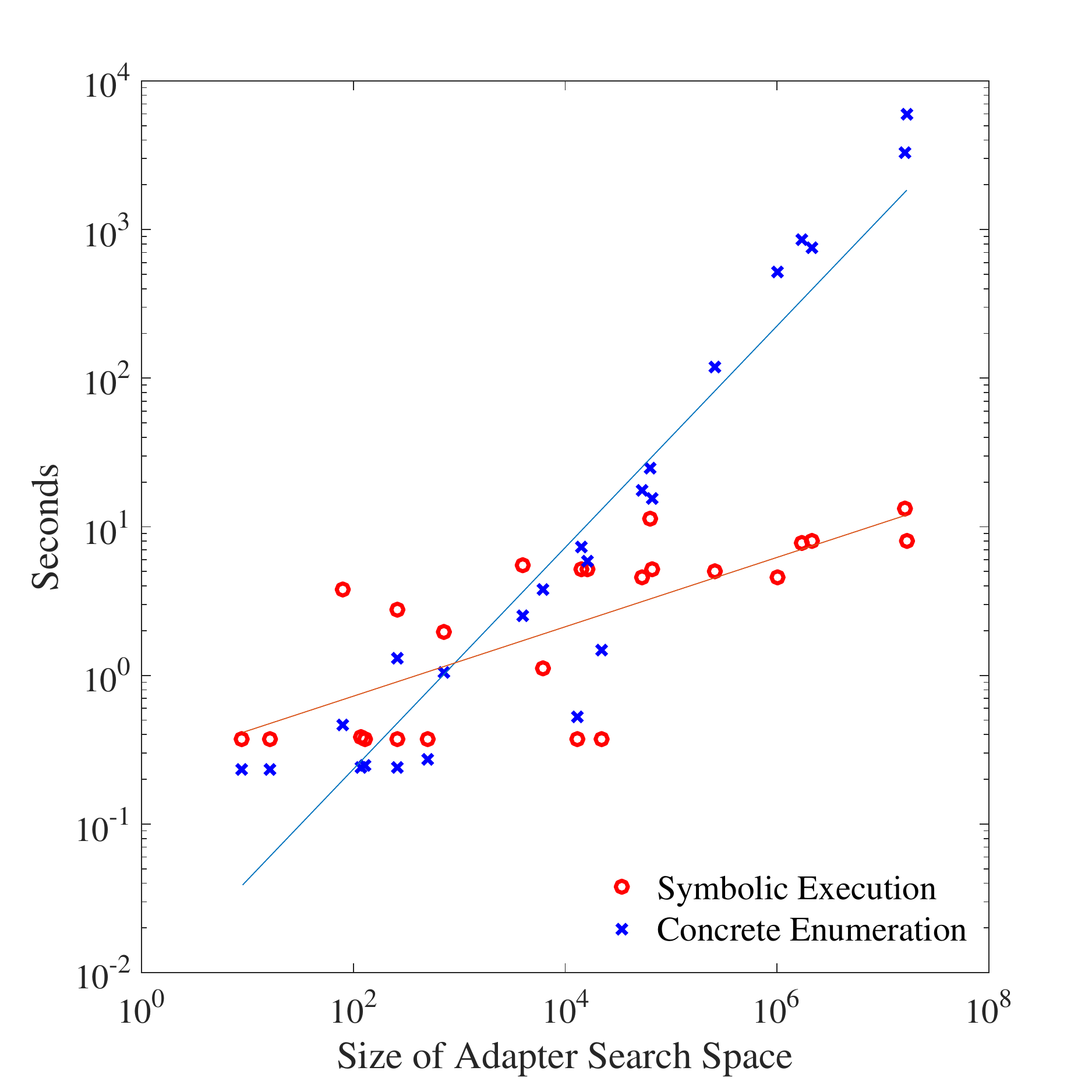}
\end{figure}

The adapter search step in our CEGIS approach need not use binary symbolic execution.
We swapped out our FuzzBALL-based adapter search step with a concrete enumeration-based adapter search.
We ensured that our concrete enumeration generated adapters, such that every adapter had the same probability of being chosen.
We synthesized every adapter, presented so far, using both adapter search implementations and captured the total adapter search time.
We also counted the size of the adapter search space for every adaptation.
In some cases, the adapter search space was too large to be concretely enumerated.
For example, the adapter search space for the \textit{killpg} $\leftarrow$ \textit{kill} adapter consists of 98.1 million arithmetic adapters.
In such cases, we reduced the size of the search space by using smaller constant bounds.
Based on the size of adapter search space, we compared the total adapter search times for both adapter search strategies.
We present the results from this comparison in Figure~\ref{fig:conc_vs_se}.
For concrete enumeration-based adapter search, Figure~\ref{fig:conc_vs_se} shows the time required to find an adapter has a consistent exponential increase with an increase in the size of adapter search space.
But, we cannot derive any such conclusion for binary symbolic execution-based adapter search.
This is because symbolic execution is more sensitive to variations in difficulty of adapter search than concrete enumeration.
We further explored this comparison between concrete enumeration and binary symbolic execution-based adapter search using an example which would allow us to control adapter search difficulty.
\begin{lstlisting}[caption={naive and SKETCH-based implementations of \textit{popCnt}}, 
label={lst:popCnt}, style=nonumbers]
unsigned short popCntNaive(unsigned short v) {
  unsigned short c;
  for (c = 0; v; v >>= 1) { c += v & 1; }
  return c;
}
unsigned short popCntSketch(unsigned short x) {
  x= (x & 0x5555)+ ((x>> 1) & 0x5555); 
  x= (x & 0x3333)+ ((x>> 2) & 0x3333); 
  x= (x & 0x0077)+ ((x>> 8) & 0x0077); 
  x= (x & 0xf)+ ((x>> 4) & 0xf); 
  return x;
}
\end{lstlisting}
The \textit{popCnt} function synthesized by SKETCH~\cite{Solar-LezamaTBSS2006} allows us to control the difficulty of adapter search.
The \textit{popCnt} function counts the number of bits set in a 16-bit value.
We present the target~(\textit{popCntNaive}) function and one variant of the inner function~(\textit{popCntSketch}) in Listing~\ref{lst:popCnt}.
The \textit{popCntSketch} function uses 8 constants~(1, 2, 4, 8, 0xf, 0x77, 0x3333, 0x5555), which can be passed as arguments instead of being hardcoded.
The argument substitution adapter family allows constant bounds to be specified to make the adapter search space finite.
By varying the constant bounds and the number of arguments (which were replaced by appropriate constants by the correct adapter) to \textit{popCntSketch}, we varied the size of the adapter search space while keeping the difficulty of adapter search uniform.
We created 24 variants of \textit{popCountSketch}.
Using each \textit{popCountSketch} variant as the inner function, and \textit{popCntNaive} as the target function, we synthesized adapters using concrete enumeration and binary symbolic execution-based adapter search.
Figure~\ref{fig:popCnt} shows the result of comparing total adapter search times across sizes of adapter search space when using concrete enumeration and binary symbolic execution-based adapter search.
Figure~\ref{fig:popCnt} shows concrete enumeration-based adapter search is faster than binary symbolic execution-based adapter search upto search spaces of size $10^3$.
But this gain quickly drops off as the size of search space approaches $10^7$.
We also created a variant of \textit{popCntSketch} that takes 6 arguments and uses them for its largest constants.
Synthesizing an adapter using this variant as the inner function creates a search space of size 3.517x$10^{18}$~(not including return value substitution adapters).
Using only binary symbolic execution-based adapter search, our tool synthesized the correct adpator in 168 seconds, with 154 seconds spent in adapter search.
Enumerating this search space concretely would take 11.15 million years.
\subsection{Reverse engineering using reference functions}
\label{sec:eval_general}
\subsubsection{Code fragment selection}:
Rockbox~\cite{rockbox} is a free replacement 3rd party firmware for digital music players.
We used a Rockbox image compiled for the iPod Nano (2g) device, based on the 32-bit ARM architecture, and disassembled it.
We dissected the firmware image into code fragments using the following rules:
(1) no code fragment could use memory, stack, floating-point, coprocessor, and supervisor call instructions,  
(2) no code fragment could branch to an address outside itself,
(3) the first instruction of a code fragment could not be conditionally executed.

The first rule disallowed code fragments from having any inputs from and outputs to memory, thereby allowing us to use the 13 general purpose registers on ARM as inputs. 
The second rule prevented a branch to an invalid address.
ARM instructions can be executed based on a condition code specified in the instruction, if the condition is not satisfied, the instruction is turned into a {\tt noop}.
The third rule disallowed the possibility of having code fragments that begin with a {\tt noop} instruction, or whose behavior depended on a condition.
The outputs of every code fragment were the last (up to) three registers written to by the code fragment.
This caused each code fragment to be used as the target code region up to three times, once for each output register. 
This procedure gave us a total of 183,653 code regions, with 61,680 of them consisting of between 3 and 20 ARM instructions.

To evaluate which code fragments can be synthesized just with our
adapter family without a contribution from a reference function, we
checked
which of these 61,680 code fragments can be adaptably substituted by a reference function that simply returns one of its arguments.
Intuitively, any code fragment that can be adaptably substituted by an
uninteresting reference function must be uninteresting itself, and so
need not be considered further.
We found 46,831 of the 61,680 code fragments could not be adaptably substituted by our simple reference function, and so we focused our further evaluation on these 46,831 code fragments that were between 3 and 20 ARM instructions long.\\
\subsubsection{Reference functions}:
Since our code fragments consisted of between 3 and 20 ARM instructions, we focused on using reference functions that can be expressed in a similar number of ARM instructions.
We used the source code of version 2.2.6 of the VLC media player~\cite{vlc} as the codebase for our reference functions.
We performed an automated search for functions that were up to 20 lines of source code.
This step gave us a total of 1647 functions.
Similar to the three rules for code fragment selection, we discarded functions that accessed memory, called other VLC-specific functions, or made system calls to find 24 reference functions. 
Other than coming from a broadly similar problem domain (media
players), our selection of reference functions was independent of the
Rockbox codebase, so we would not expect that every reference function
would be found in Rockbox.

\subsubsection{Results}
We used the type conversion adapter family along with the return value
substitution family, disallowing return value substitution adapters
from setting the return value to be a type-converted argument of the
reference function (which would lead to uninteresting adapters).
But we allowed the reference function arguments to be replaced by
unrestricted 32-bit constants, and we assumed each code segment takes
up to 13 arguments.
The size of this adapter search space can be calculated using the following formula:\\
$8 * \sum_{k=0}^{k=13} (2^{32})^{13-k} * \comb{13}{k} * \perm{13}{k} * 8^k$ \\
The first multiplicative factor of 8 is due to the 8 possible return value substitution adapters. 
The permutation and combination operators occur due to the choices of arguments for the target code fragment and reference functions~(we assumed both have 13 arguments).
The final $8^k$ represents the 8 possible type conversion operators that a type conversion adapter can apply.
The dominant factor for the size of the adapter search space comes from size of the set of possible constants.
Our adapter family used unrestricted 32-bit constants, leading to a constants set of size $2^{32}$.

With this adapter family set up, we ran adapter synthesis trying to adaptably substitute each of the 46,831 code fragments by each reference function .
This gave us a total of 1,123,944~(46831*24) adapter synthesis tasks, with each adapter synthesis search space consisting of 1.353 x $10^{127}$ adapters, too large for concrete enumeration.
%
%This calculation is using 10000 adapters concretely enumerated per second, which is a loose upper bound on what we observed when doing concrete adapter enumeration for popcnt.
%
We set a 5 minute hard time limit and a total memory limit of 2 GB per adapter synthesis task.
We split the adapter synthesis tasks with each reference function into 32 parallel jobs creating a total of 768~(32*24) parallel jobs.
We ran our jobs on a machine cluster running CentOS 6.8 with 64-bit Linux kernel version 2.6.32 and Intel Xeon E5-2680v3 processors.
We present our results in Table~\ref{table:general}.
The full set of results is presented in Section~\ref{sec:all_tables} of
the Appendix.
\input{revengg_general}
The first column shows the reference functions chosen from the VLC media player source code.
The \textit{\#(full)} column reports how many code fragments were found to be adaptably substitutable~(represented by the value for \textit{\#}), and how many of those exploited the full generality of the reference function~(represented by the value of \textit{full}). 
We report average number of steps, average total running time (average solver time), average total time spent in adapter search steps (average time during the last adapter search step) in columns \textit{steps}, \textit{total time (solver)}, \textit{AS time (last)} respectively. 
In case of timeouts, only average solver time is reported since the average total running time was 5 minutes.\\
\subsubsection{Clustering using random tests:} For every reference function, we can either have a conclusion that finds an adapter, or finds the fragment to not be adaptably substitutable, or runs out of time.
Our adapter synthesis tool finds adaptable substitution using 18 out of the 24 reference functions.
For every reference function, we cluster its adapted versions using 100000 random tests: all adapted versions of a reference function that report the same output for all inputs are placed in the same cluster.
The number of clusters is reported in the \textit{\#clusters} column.
For each reference function, we then manually examine these clusters to judge which adapted versions use the complete functionality of that reference function; these are the cases where describing the functionality of the target fragment in terms of the reference function is mostly likely to be concise and helpful.
This took us less than a minute of manual effort for each reference function because we understood the intended semantic functionality of every reference function~(we had its source code).
We found several instances of adapters using the full generality of the reference function for 11 reference functions.
%
%It must be noted that our choice of reference functions was not based on their occurrence in the Rockbox firmware image.
%
\begin{figure}
\centering
\includegraphics[width=\columnwidth]{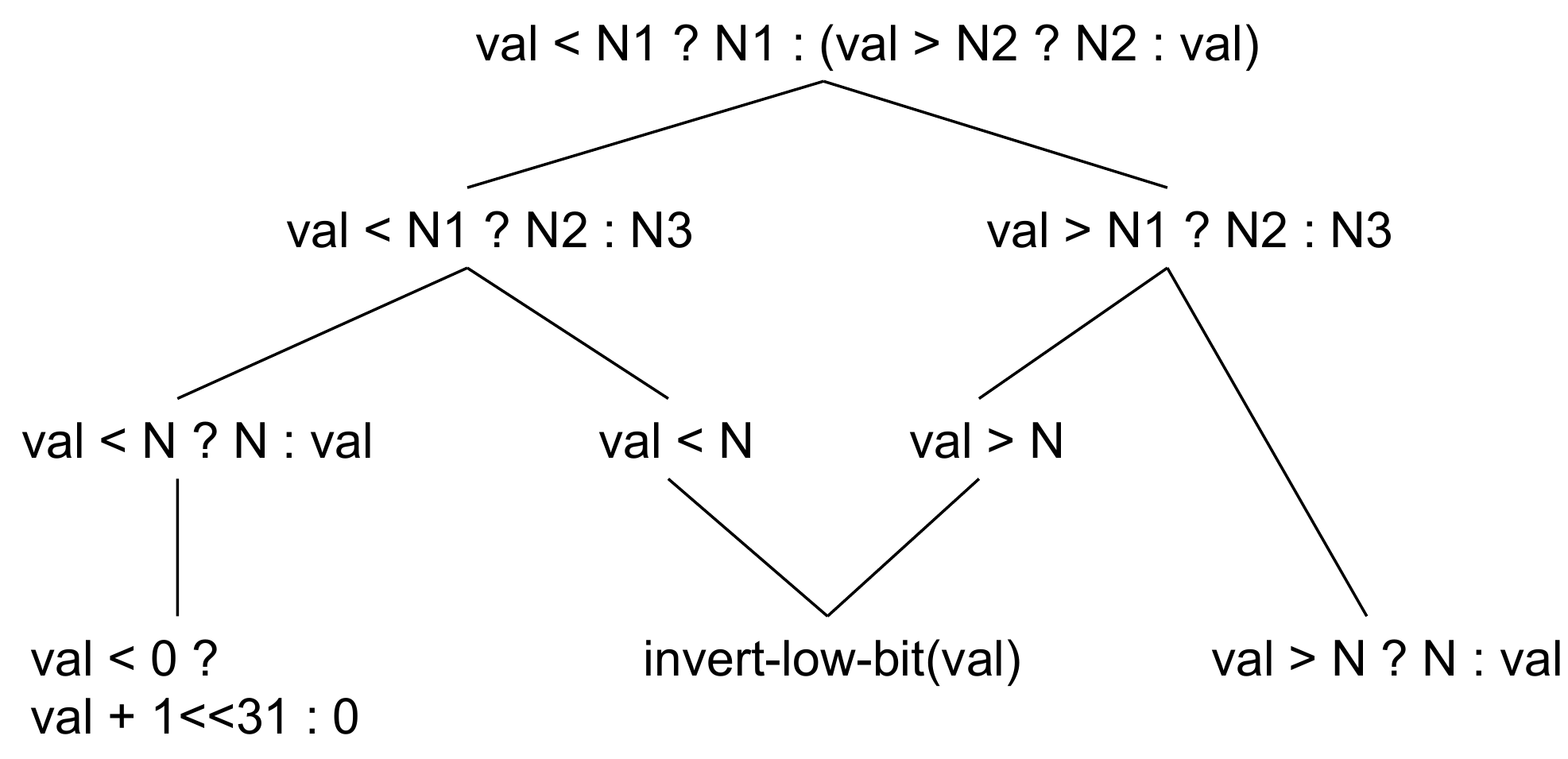}
\caption{Subset of partial order relationship among adapted clamp instances}
\label{fig:clamp_partial_order}
\end{figure}

We found that a majority of our found adapters exploit specific functionality of the reference functions.
We explored this observation further by manually summarizing the semantics of the 683 adapters reported for {\tt clamp}.
We found that these 683 adapters have a partial order between them created by our adapter families of type conversion and return value substitution.
We present a subset of this partial order as a lattice-like diagram in Figure~\ref{fig:clamp_partial_order}.
To explain one unexpected example, the {\tt invert-low-bit} operation on a value {\tt v} can be implemented in terms of {\tt val < N} by setting {\tt val} to the low bit of {\tt v} zero-extended to 32 bits and {\tt N} to 1, and zero-extending the low 1 bit of the return value of {\tt val < N} to 32 bits.
Some such functionalities owe more to the flexibility of the adapter
family than they do to the reference function.
These results suggest it would be worthwhile in the future to prune
them earlier by searching for instances of the simplest reference
functions first, and then excluding these from future searches.

\begin{figure}[ht]
\centering
\includegraphics[width=0.5\textwidth]{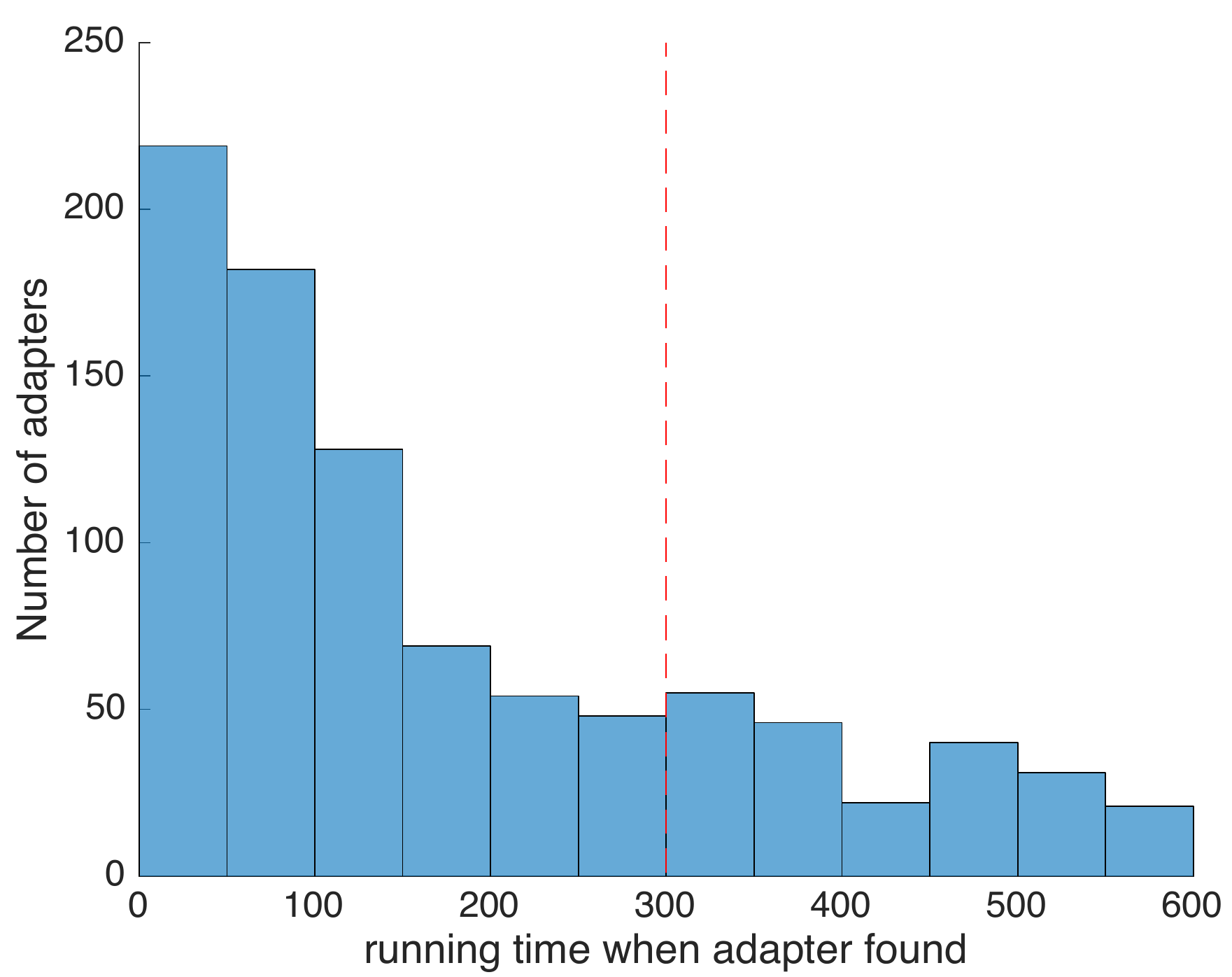}
\caption{Running times for synthesized adapters using {\tt clamp} reference function}
\label{fig:clamp_hist}
\end{figure}

\begin{figure}[ht]
\centering
\includegraphics[width=0.5\textwidth]{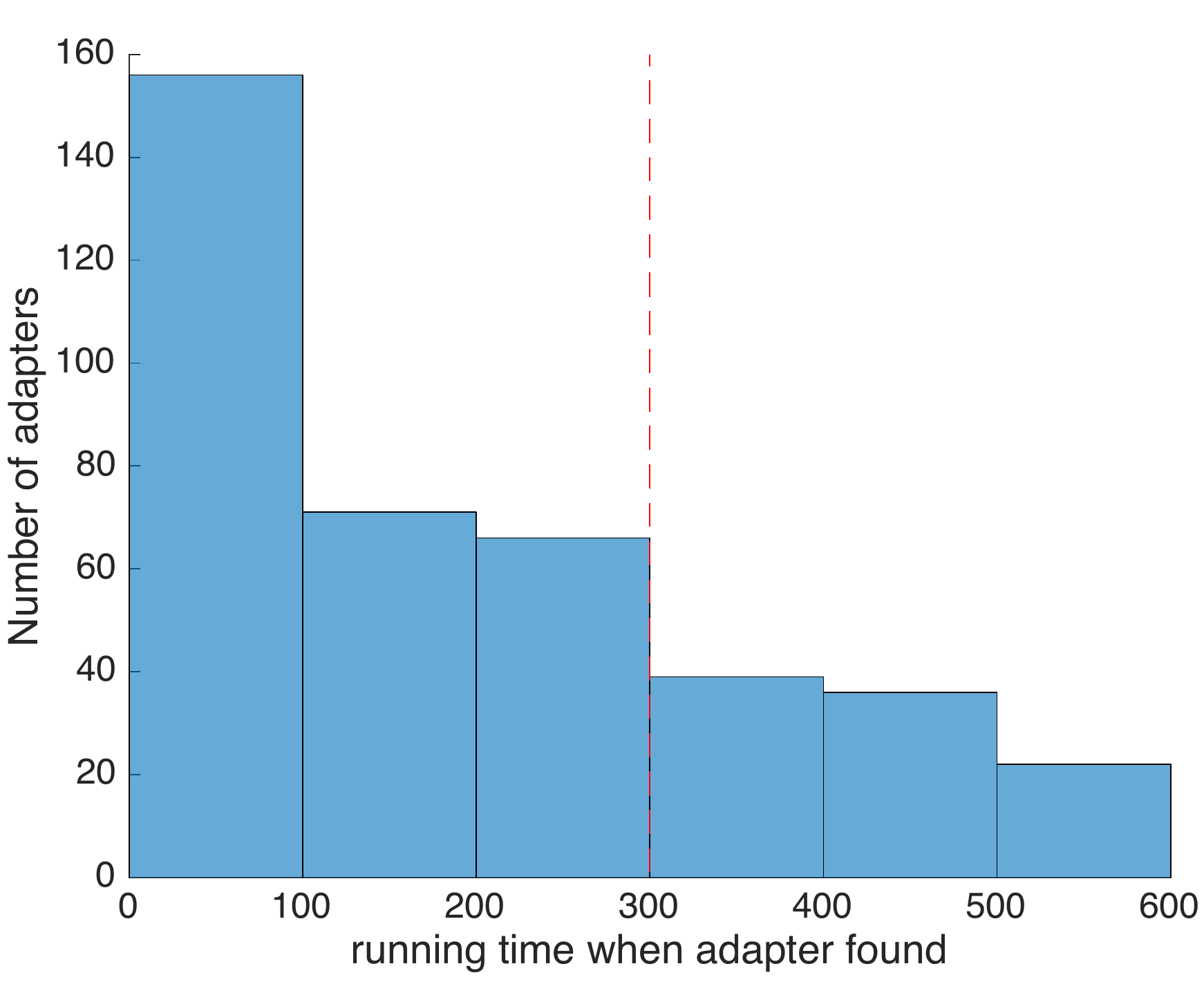}
\caption{Running times for synthesized adapters using {\tt median} reference function}
\label{fig:median_hist}
\end{figure}

\begin{figure}
\centering
\includegraphics[width=0.5\textwidth]{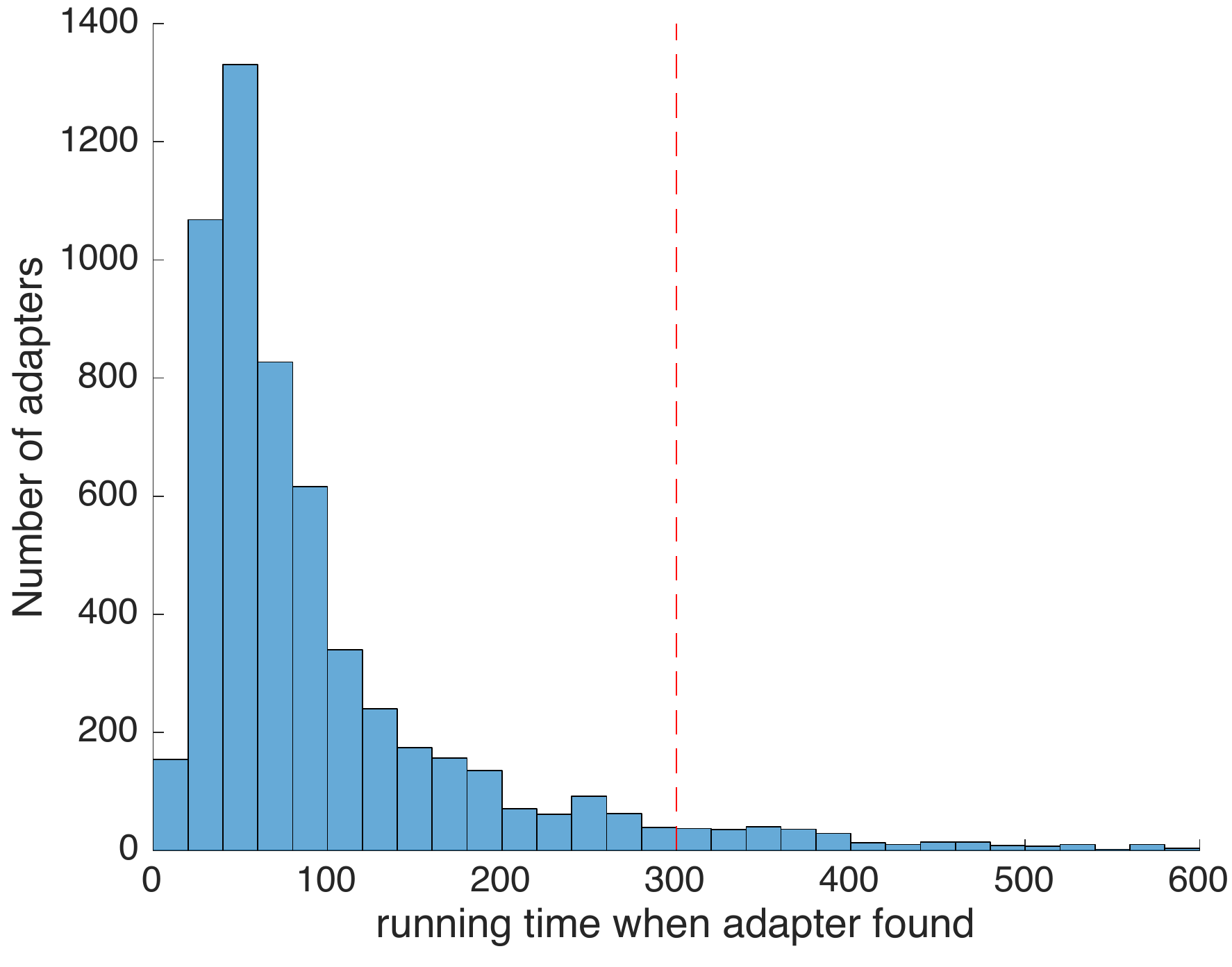}
\caption{Running times for synthesized adapters using {\tt tile\_pos} reference function}
\label{fig:tilepos_hist}
\end{figure}

Timeouts were the third conclusion of each adapter synthesis task as reported in Table~\ref{table:general}.
We report a histogram of the total running time used to find adapters in Figure~\ref{fig:tilepos_hist} for the {\tt tile\_pos} reference function, which had the most timeouts. 
Similar histograms for {\tt clamp} and {\tt median} reference function are reported in Figures~\ref{fig:clamp_hist},~\ref{fig:median_hist}.

The number of adapters found after 300 seconds decreases rapidly, consistent with the mean total running time~(subcolumn \textit{total time (solver)} under column \textit{adapter} in Table~\ref{table:general}) of 53.5 seconds for the {\tt tile\_pos} reference function.
Table~\ref{table:general} also shows that the total running time, when our tool concludes with finding an adapter, is significantly lesser than 300 seconds for all reference functions that reported adapters. 
Though setting any finite timeout can cause some instances to be lost,
these results suggest that a 300-second timeout was appropriate for
this experiment, and that most timeouts would not have led to adapters.
\subsection{Comparing adapter families}
\label{sec:eval_compare}
We also explored the tradeoff between adapter search space size and effectiveness of the adapter family.
We ran all 46,831 target code fragments with {\tt clamp} as the reference function using two additional adapter families beyond the combination of type conversion family with return value substitution described above.
The first adapter family allowed only argument permutation and the second allowed argument permututation along with substitution with unrestricted 32-bit constants.
We ran the first adapter family setup (argument permutation + return value substitution) with a 2.5 minute hard time limit, the second adapter family setup (argument substitution + return value substitution) with a 5 minute hard time limit, and the third adapter family setup (argument substitution + return value substitution) was the same as the previous subsection with also a 5 minute hard time limit. 
We present our results in Table \ref{table:compare}.
\begin{table}
\centering
\caption{Comparing adapter families with 46,831 target code fragments and {\tt clamp} reference function}
\label{table:compare}
\begin{tabular}{|l|l|l|l|l|}
\hline
                                                                   & size           & \#-ad & \#-inequiv & \#-timeout \\ \hline
\begin{tabular}[c]{@{}l@{}}arg\_perm+\\ ret\_sub-2.5m\end{tabular} & 4.98E+10       & 9     & 46803      & 19         \\ \hline
\begin{tabular}[c]{@{}l@{}}arg\_sub+\\ ret\_sub-2.5m\end{tabular}  & 1.3538427E+126 & 705   & 45782      & 344        \\ \hline
\begin{tabular}[c]{@{}l@{}}type\_conv+\\ ret\_sub-5m\end{tabular}  & 1.3538430E+126 & 683   & 40553      & 5595       \\ \hline
\end{tabular}
\end{table}
As expected, the number of timeouts increases with an increase in the size of adapter search space. 
Table \ref{table:compare} also shows that, for {\tt clamp}, a simpler adapter family is better at finding adapters than a more expressive family, because more searches can complete within the timeout.
But, this may not be true for all reference functions.
Table~\ref{table:compare} suggests that, when computationally feasible, adapter families should be tried in increasing order of expressiveness to have the fewest timeouts overall.
We plan to explore this tradeoff between expressiveness and effectiveness of adapter families in the future.

%% file: adapter_results_1.tex
\begin{table}[t]
\centering
\caption{adapter Synthesis over 13130 function pairs without memory-based equivalence checking}
\label{table:libc-evaluation}
\begin{tabular}{|l|l|l|l|l|l|}
\hline
\begin{tabular}[c]{@{}l@{}}adapter \\ type\end{tabular}     &
Inequiv. & \begin{tabular}[c]{@{}l@{}}adapters\\ Found\end{tabular}
& Timeout & \begin{tabular}[c]{@{}l@{}}Target \\ function\\
crashed\end{tabular} \\ \hline
\begin{tabular}[c]{@{}l@{}} arg. sub.\end{tabular} & 8887
& 382 & 3014 &
847 \\
\hline
type conv.  & 8909
& 383 & 2989 &
849 \\
\hline
\end{tabular}
\end{table}

%the numbers in the above table's Timeout column include iteration limit timeouts 
%these were derived using the check_iterlimit.cpp file at MSI

%we reduce the number of inequivalences by 34 because that is the number of function pairs
% for which the words "after too many" was found in the log of that function pair
% but near the end of that function pair's log, we found them to be inequivalent

%similar logic was applied in check_iterlimit.cpp for function pairs with "missing results from check run" and for function pairs for which a "Final adapter" was reported

%finally, adding the search string "Solver died with result code 134" instead of "after too many" allowed us to search for number of solver timeouts

%the final numbers in the table are harvest_results[ineq|missing|final] - `grep "ineq: |missing results: | final: " results/<simple-1 or typeconv-4>/iterlimit_solvertimeout.txt` for the 3 columns, ineq, missing results, final adapter. The total of these 3 subtractions was added to the Timeout column. 

\begin{table}[]
\caption{adapters found within eglibc-2.19}
\label{table:libc-adapters}
\resizebox{\columnwidth}{!}{%
\begin{tabular}{ll}
\hline
$f_1$ $\leftarrow$ $f_2$ or $f_1$ $\leftrightarrow$ $f_2$                                                                                                                                                                      & adapter                                                                                   \\ \hline
\begin{tabular}[c]{@{}l@{}}abs(1) $\leftarrow$ labs(1)\\ abs(1) $\leftarrow$ llabs(1)\end{tabular}                                                                                                                             & \begin{tabular}[c]{@{}l@{}}32-to-64S(\#0) and \\ 32-to-64Z(return value)\end{tabular}     \\ \hline
\begin{tabular}[c]{@{}l@{}}labs(1) $\leftrightarrow$ llabs(1)\end{tabular}                                                                                                                           & \#0                                                                                       \\ \hline
\begin{tabular}[c]{@{}l@{}}ldiv(1) $\leftrightarrow$ lldiv(1)\end{tabular}                                                                                                                           & \#0                                                                                       \\ \hline
\begin{tabular}[c]{@{}l@{}}ffs(1) $\leftarrow$ ffsl(1)\\ ffs(1) $\leftarrow$ ffsll(1)\end{tabular}                                                                                                                              & 32-to-64S(\#0)                                                                            \\ \hline
\begin{tabular}[c]{@{}l@{}}ffsl(1) $\leftrightarrow$ ffsll(1)\end{tabular}                                                                                                                           & \#0                                                                                       \\ \hline
setpgrp(0) $\leftarrow$ setpgid(2)                                                                                                                                                                            & 0, 0                                                                                    \\ \hline
wait(1) $\leftarrow$ waitpid(3)                                                                                                                                                                               & -1, \#0, 0                                                                               \\ \hline
wait(1) $\leftarrow$ wait4(4)                                                                                                                                                                                 & -1, \#0, 0, 0                                                                          \\ \hline
waitpid(3) $\leftarrow$ wait4(4)                                                                                                                                                                              & \#0, \#1, \#2, 0                                                                        \\ \hline
wait(1) $\leftarrow$ wait3(3)                                                                                                                                                                                 & \#0, 0, 0                                                                               \\ \hline
wait3(3) $\leftarrow$ wait4(4)                                                                                                                                                                                & -1, \#0, \#1, \#2                                                                       \\ \hline
umount(1) $\leftarrow$ umount2(2)                                                                                                                                                                             & \#0, 0                                                                                  \\ \hline
\begin{tabular}[c]{@{}l@{}}putchar(1) $\leftrightarrow$ putchar\_unlocked\\ putwchar(1) $\leftrightarrow$ putwchar\_unlocked(1) \end{tabular} & \#0                                                                                       \\ \hline
\begin{tabular}[c]{@{}l@{}}recv(4) $\leftarrow$ recvfrom(6)\\ send(4) $\leftarrow$ sendto(6)\end{tabular}                                                                                                            & \begin{tabular}[c]{@{}l@{}}32-to-64S(\#0), \#1, \#2, \\ 32-to-64S(\#3), 0, 0\end{tabular} \\ \hline
\begin{tabular}[c]{@{}l@{}}atol(1) $\leftrightarrow$ atoll(1)\end{tabular}                                                                                                            & \begin{tabular}[c]{@{}l@{}}\#0\end{tabular} \\ \hline
\begin{tabular}[c]{@{}l@{}}atol(1) $\leftarrow$ strtol(3)\\ atoi(1) $\leftarrow$ strtol(3) \\ atoll(1) $\leftarrow$ strtoll(3) \end{tabular}                                                           & \begin{tabular}[c]{@{}l@{}}\#0, 0, 10 \end{tabular} \\ \hline \hline
isupper(1) $\leftarrow$ islower(1)                                                                                                                                                                            & \#0 + 32                                                                                  \\ \hline
islower(1) $\leftarrow$ isupper(1)                                                                                                                                                                            & \#0 - 32                                                                                  \\ \hline
killpg(1) $\leftarrow$ kill(1)                                                                                                                                                                            & -\#0, \#1                                                                                  \\ \hline
\end{tabular}
}
\end{table}

%% file: revengg_general.tex
\begin{table*}
	\centering
\caption{Reverse engineering results using 46831 target code fragments from a Rockbox firmware image and 24 reference functions from VLC media player grouped by the three overall possible terminations of adapter synthesis. The \textit{\#(full)} column reports how many code fragments were found to be adaptably substitutable, and how many of those exploited the full generality of the reference function.}
\label{table:general}
\begin{tabular}{@{}c@{}c@{\hspace{0.2mm}}c@{\hspace{0.2mm}}cccccccccc@{}}
\toprule
		 & \multicolumn{5}{c}{adapter} & \multicolumn{3}{c}{not
	substitutable} & \multicolumn{4}{c}{timeout} \\ \midrule
		\multicolumn{1}{c}{fn\_name} & \#(full) & \#cluster & steps &
	\begin{tabular}[c]{@{}c@{}}total \\ time \\ (solver)\end{tabular} &
	\begin{tabular}[c]{@{}c@{}}AS\\ time\\ (last)\end{tabular} & \# &
		steps & \begin{tabular}[c]{@{}c@{}}total \\ time \\
	(solver)\end{tabular} & \# & steps &
	\begin{tabular}[c]{@{}c@{}}solver\\ time\end{tabular} &
	\begin{tabular}[c]{@{}c@{}}stops\\ (AS/CE)\end{tabular} \\ \midrule
		\multicolumn{1}{c}{clamp} & 683(177) & 110 & 12.9 & 99.3(12.1) &
		82.2(32.5) & 40553 & 7.7 & 63.0(6.4) & 5595 & 16.5 & 44.3 & 5416/179
		\\ \midrule
		\multicolumn{1}{c}{prev\_pow\_2} & 32(0) & 6 & 4.7 & 6.1(0.3) &
		1.8(0.9) & 46767 & 4.3 & 7.5(0.5) & 32 & 1 & 289.4 & 0/32 \\
		\midrule
		\multicolumn{1}{c}{abs\_diff} & 575(5) & 75 & 10.5 & 20.0(1.3) &
		7.0(1.8) & 46250 & 8.2 & 18.7(1.4) & 6 & 5.7 & 286.5 & 0/6 \\
		\midrule
		\multicolumn{1}{c}{bswap32} & 115(8) & 19 & 8.7 & 16.6(1.2) &
		4.3(1.2) & 46708 & 4.7 & 8.2(0.5) & 8 & 2.8 & 293.5 & 0/8 \\
		\midrule
		\multicolumn{1}{c}{integer\_cmp} & 93(5) & 15 & 9.6 & 21.4(2.2) &
		12.6(4.7) & 46467 & 5.2 & 15.3(1.8) & 271 & 3.1 & 247.2 & 3/268 \\
		\midrule
		\multicolumn{1}{c}{even} & 3(2) & 3 & 5.7 & 11.3(0.6) & 4.3(2.3) &
		46823 & 4.2 & 12.7(0.9) & 5 & 1.8 & 116.5 & 0/5 \\ \midrule
		\multicolumn{1}{c}{div255} & 4(0) & 2 & 5 & 6.5(0.3) & 2.5(1.5) &
		46823 & 4.4 & 7.6(0.5) & 4 & 2.5 & 294.2 & 0/4 \\ \midrule
		\multicolumn{1}{c}{reverse\_bits} & 276(0) & 11 & 9 & 25.3(2.9) &
		9.1(1.9) & 46541 & 12.5 & 50.9(5.6) & 14 & 3 & 292.3 & 0/14 \\
		\midrule
		\multicolumn{1}{c}{binary\_log} & 48(0) & 5 & 6.7 & 23.6(5.9) &
		12.6(8.8) & 46528 & 4 & 25.6(6.4) & 255 & 1.2 & 207.3 & 19/236 \\
		\midrule
		\multicolumn{1}{c}{median} & 332(42) & 60 & 13.7 & 119.2(26.7) &
		101.4(33.9) & 32171 & 6.5 & 89.8(15.1) & 14328 & 13.6 & 65.4 &
		14184/144 \\ \midrule
		\multicolumn{1}{c}{hex\_value} & 0 & 0 & 0 & 0) & 0 & 46354 & 3.2
		& 9.2(2.1) & 477 & 1 & 268.8 & 2/475 \\ \midrule
		\multicolumn{1}{c}{get\_descriptor\_len} & 22(9) & 2 & 9 &
		16.7(0.6) & 4.8(1.6) & 46625 & 5.4 & 11.7(0.7) & 184 & 1.4 & 233.4 &
		0/184 \\ \midrule
		\multicolumn{1}{c}{tile\_pos} & \hspace{0.2mm} 5617(407) & 909 & 10.9 &
		53.5(23.1) & 42.5(18.1) & 24696 & 8 & 67.6(27.1) & 16518 & 7.6 &
		280.5 & 16441/77 \\ \midrule
		\multicolumn{1}{c}{diract\_pic\_n\_bef\_m} & 330(2) & 18 & 13.2 &
		25.7 (3.0) & 12.7(2.1) & 46393 & 6.6 & 15.3(1.3) & 108 & 51.5 & 138
		& 74/34 \\ \midrule
		\multicolumn{1}{c}{ps\_id\_to\_tk} & 0 & 0 & 0 & 0 & 0 & 46721 &
		4.4 & 15.8(2.4) & 110 & 1.1 & 258.8 & 3/107 \\ \midrule
		\multicolumn{1}{c}{clz} & 41(0) & 7 & 18.6 & 39.0(4.5) & 16.2(2.5)
		& 46727 & 7.8 & 16.7(2.1) & 63 & 5.1 & 143.6 & 1/62 \\ \midrule
		\multicolumn{1}{c}{ctz} & 46(0) & 4 & 5.9 & 16.2(3.8) & 7.1(3.5) &
		46701 & 3.4 & 19.5(6.2) & 84 & 1.5 & 283.7 & 4/80 \\ \midrule
		\multicolumn{1}{c}{popcnt\_32} & 0 & 0 & 0 & 0 & 0 & 46802 & 5.6 &
		11.5(0.8) & 29 & 1 & 295.4 & 0/29 \\ \midrule
		\multicolumn{1}{c}{parity} & 0 & 0 & 0 & 0 & 0 & 46821 & 5 &
		10.0(0.6) & 10 & 1.2 & 266.3 & 0/10 \\ \midrule
		\multicolumn{1}{c}{dv\_audio\_12\_to\_16} & 0 & 0 & 0 & 0 & 0 &
		46637 & 3.9 & 17.7(2.8) & 194 & 1 & 290.3 & 1/193 \\ \midrule
		\multicolumn{1}{c}{is\_power\_2} & 0 & 0 & 0 & 0 & 0 & 46801 & 3.8
		& 9.1(1.4) & 30 & 3.7 & 291.1 & 0/30 \\ \midrule
		\multicolumn{1}{c}{RenderRGB} & 763(2) & 64 & 10.8 & 27.5(1.5) &
		10.4(2.8) & 46061 & 5.7 & 17(0.9) & 7 & 4.4 & 115.7 & 0/7 \\
		\midrule
		\multicolumn{1}{c}{decode\_BCD} & 0 & 0 & 0 & 0 & 0 & 46824 & 4.7
		& 8.8(1.1) & 7 & 1.9 & 126.1 & 0/7 \\ \midrule
		\multicolumn{1}{c}{\begin{tabular}[c]{@{}c@{}}mpga\_get\_\\
		frame\_samples\end{tabular}} & 22(15) & 4 & 5 & 7.9(0.9) & 2.6(1.8)
				& 46235 & 3.4 & 9.3(2.1) & 574 & 1 & 289.5 & 4/570 \\
				\bottomrule
			\end{tabular}
		\end{table*}

%% file: discussion.tex
\section{Limitations and Future Work}\label{sec:discussion}

We represented our synthesized adapters by an assignment of concrete values to symbolic variables and manually checked them for correctness. 
adapters can be automatically translated into binary code which can replace the original function with the adapter function.
We plan to automate generation of such adapter code in the future.

During every adapter search step, symbolic execution explores all feasible paths, including paths terminated on a previous adapter search step because they did not lead to a correct adapter.
Once an adapter is found, the next adapter search can be accelerated by saving the state of adapter search, and picking up symbolic execution from the last path that led to a correct adapter in the next adapter search.

Our tool currently presumes that all behaviors of the target function must be matched, modulo failures such as null dereferences.
Using a tool like Daikon~\cite{ernst2007daikon} to infer the preconditions of a function from its uses could help our tool find adapters that are correct for correct uses of functions, such as {\em isupper} and {\em islower}.

adapter synthesis requires us to find if {\em there exists} an adapter such that {\em for all} inputs to the target function, the output of the target function and the output of the adapted inner function are equal. 
Thus the synthesis problem can be posed as a single query whose variables have this pattern of quantification (whereas CEGIS uses only quantifier-free queries).
We plan to explore using solvers for this $\exists\forall$ fragment of first-order bitvector logic, such as Yices~\cite{yices}.

Symbolic execution can only check equivalence over inputs of bounded
size, though improvements such as path
merging~\cite{KuznetsovKBC2012,AvgerinosRCB2014} can improve scaling.
Our approach could also integrate with any other equivalence checking
approach that produces counterexamples, including ones that
synthesize inductive invariants to cover unbounded
inputs~\cite{SrivastavaG2009}, though we are not aware of any
existing binary-level implementations that would be suitable.

%% file: related_work.tex
\section{Related Work}\label{sec:related-work}
\subsection{Detecting Equivalent Code}
The majority of previous work in this area has focused on detecting \emph{syntactically} equivalent code, or `clones,' which are, for instance, the result of copy-and-paste \cite{Kamiya:2002:CMT:636188.636191,Li:2004:CTF:1251254.1251274,Jiang:2007:DSA:1248820.1248843}. 
%
%Applications of detecting functionally equivalent code (aside from our motivating application of multivariant execution) include functionality-based refactoring, semantic-aware code search, and checking `yesterday's code against today's.'  
%
Jiang et al.~\cite{Jiang:2009:AMF:1572272.1572283} propose an algorithm for automatically detecting functionally equivalent code fragments using random testing and allow for limited types of adapter functions over code inputs --- specifically permutations and combinations of multiple inputs into a single struct. 
%Similar to our work, both \cite{Jiang:2009:AMF:1572272.1572283} and \cite{Ramos:2011:PLE:2032305.2032360} define functional equivalence based on input and output behavior. 
%The key difference between our approach and \cite{Jiang:2009:AMF:1572272.1572283} is that we rely on symbolic execution as opposed to random testing, and that we allow for more interesting adapter functions over code inputs. 
Ramos et al.~\cite{Ramos:2011:PLE:2032305.2032360} present a tool that checks for equivalence between arbitrary C functions using symbolic execution.
While our definition of functional equivalence is similar to that used by Jiang et al. and Ramos et al., our adapter families capture a larger set of allowed transformations during adapter synthesis than both. 

%Detecting pieces of equivalent code is useful for many applications including refactoring, code understanding, optimization, and plagiarism detection. 
%
Amidon et al.~\cite{program_fracture} describe a technique for fracturing a program into pieces which can be replaced by more optimized code from multiple applications.
They mention the need for automatic generation of adapters which enable replacement of pieces of code which are not immediately compatible.
While Amidon et al. describe a parameter reordering adapter, they do not mention how automation of synthesis of such adapters can be achieved.
David et al.~\cite{statistical_similarity} decompose binary code into smaller pieces, find semantic similarity between pieces, and use statistical reasoning to compose similarity between procedures.
Since this approach relies on pieces of binary code, they cannot examine binary code pieces that make function calls and check for semantic similarity across wrappers around function calls.
Goffi et al.~\cite{goffi} synthesize a sequence of functions that are equivalent to another function w.r.t a set of execution scenarios. 
Their implementation is similar to our concrete enumeration-based adapter search which produces equivalence w.r.t. a set of tests.
In the hardware domain, adapter synthesis has been applied to low-level combinatorial circuits by Gasc\'{o}n et al~\cite{gascon}.
They apply equivalence checking to functional descriptions of a low-level combinatorial circuit and reference implementations while synthesizing a correct mapping of the input and output signals and setting of control signals. 
They convert this mapping problem into a exists/forall problem which is solved using the Yices SMT solver~\cite{yices}. 
%
% Their permutation of input signals and output signals are similar to our argument substitution and return value adapters.
%
%However, their technique depends on the user specifying input and control signals for reference implementations whereas our technique does not depend on any such prior classification. 
%
%\subsection{Variant Generation}
%
%Another similar approach to developing variants relies on compiler-based randomization~\cite{Larsen:2014:SAS:2650286.2650803}. 
%
%It can be convenient to modify a compiler to support randomization because compilers already have support for many of the analyses required for randomization and are set up to target many different architectures. 
%
%However, compiler-based variant generation requires that the source code of the program to be randomized is available and that it is possible to customize the compiler. 
%
%Because we check for functional equivalence at the binary level, our approach does not require source code and is compatible with proprietary compilers. 
%
%Compiler-based approaches to diversity are also limited in the types of diversity they can introduce. 
%
\subsection{Component Retrieval}
Type-based component retrieval was an active area of research in the past.
Many papers in this area~\cite{rittri},~\cite{runciman1989},~\cite{runciman1991} focused on the problem of finding a function, whose polymorphic type is known to the user, within a library of software components.
Type-based hot swapping~\cite{duggan} and signature matching~\cite{zaremski} were also active areas of related research in the past.
These techniques relied on adapter-like operations such as currying or uncurrying functions, reordering tuples, and type conversion.
Reordering, insertion, deletion, and type conversion are only some of the many operations supported by our adapters. 
These techniques can only be applied at the source level, whereas our adapter synthesis technique can be applied at source and binary levels

\subsection{Component Adaptation}
Component adaptation was another related active area of research in the past.
This includes techniques for adapter specification~\cite{nimble}, for component adaptation using formal specifications of components~\cite{spartacus},~\cite{penix},~\cite{penix1995},~\cite{yellin},~\cite{bracciali}.
Component adaptation has also been performed at the Java bytecode level~\cite{keller}, as well as the C bitcode level~\cite{nita}.
Behavior sampling~\cite{podgurski} is a similar area of research for finding equivalence over a small set of input samples.
However, these techniques either relied on having a formal specification of the behavior of all components in the library to be searched, or provided techniques for translating a formally specified adapter~\cite{nimble}.

\subsection{Program Synthesis}
Program synthesis is an active area of research that has many applications including generating optimal instruction sequences \cite{Massalin:1987:SLS:36206.36194,Joshi:2002:DGS:512529.512566}, automating repetitive programming, filling in low-level program details after programmer intent has been expressed \cite{Solar-LezamaTBSS2006}, and even binary diversification \cite{Jacob2008}. 
Programs can be synthesized from formal specifications \cite{Manna:1980:DAP:357084.357090}, simpler (likely less efficient) programs that have the desired behavior \cite{Massalin:1987:SLS:36206.36194,Solar-LezamaTBSS2006,Joshi:2002:DGS:512529.512566}, or input/output oracles \cite{Jha:2010:OCP:1806799.1806833}. 
We take the second approach to specification, treating existing functions as specifications when synthesizing adapter functions.

%% file: conclusion.tex
\section{Conclusion}\label{sec:conclusion}
We presented a new technique to search for semantically-equivalent pieces of code which can be substituted while adapting differences in their interfaces.
This approach is implemented at the binary level, thereby enabling wider applications and consideration of exact run-time behavior.
We implemented adapter synthesis for x86-64 and ARM binary code.
We presented examples demonstrating applications towards security, deobfuscation, efficiency, and library replacement, and an evaluation using the C library.
Our adapter families can be combined to find sophisticated adapters as shown by adaptation of RC4 implementations.
While finding thousands of functions to not be equivalent, our tool reported many instances of semantic equivalence, including C library functions such as \textit{ffs} and \textit{ffsl}, which have assembly language implementations.
Our comparison of concrete enumeration-based adapter search with binary symbolic execution-based adapter search allows users of adapter synthesis to choose between the two approaches based on size of adapter search space.
We selected more than 61,000 target code fragments from a 3rd party firmware image for the iPod Nano 2g and 24 reference functions from the VLC media player. 
Given a adapter search space of 1.353 x $10^{127}$ adapters, we used binary symbolic execution-based adapter search to run more than a million adapter synthesis tasks.
Our tool finds dozens of instances of several reference functions in the firmware image, and confirms that the process of understanding the semantics of binary code fragments can be automated using adapter synthesis.
Our results show that the CEGIS approach for adapter synthesis of binary code is feasible and sheds new light on potential applications such as searching for efficient clones, deobfuscation, program understanding, and security through diversity.

%% file: new_appendix.tex
\section{Appendix}
\label{sec:appendix}

\subsection{Reverse engineering expanded tables}
\label{sec:all_tables}
For the results reported in Section~\ref{sec:eval_general}, we report detailed metrics for the three possible conclusions, adapter
found, not substitutable, timed out, in the
Tables~\ref{table:adapters_full},~\ref{table:inequiv_full},~\ref{table:timeouts_full}
respectively.
The \textit{AS-stops/CE-stops} column in Table~\ref{table:timeouts_full} reports the number of times a timeout resulted in an adapter search step or counter-example search step to be halted.
In the first column, after each reference function\rq s name, the {\tt
\#N} within parenthesis reports the number of arguments taken by the reference function.
\input{adapters_full_table}
\input{inequiv_full_table}
\input{timeouts_full_table}

%% file: adapters_full_table.tex
% % \usepackage{booktabs}
\begin{table*}[]
\centering
\caption{Metrics for adapters for all reference functions}
\label{table:adapters_full}
	\begin{tabular}{@{}llllllllll@{}}
		\toprule
		fn\_name & \# & \#full & \#clusters & steps &
	\begin{tabular}[c]{@{}l@{}}total time \\ (solver)\end{tabular} &
	\begin{tabular}[c]{@{}l@{}}CE total time \\ (solver)\end{tabular} &
	\begin{tabular}[c]{@{}l@{}}CE last time \\ (solver)\end{tabular} &
	\begin{tabular}[c]{@{}l@{}}AS total time \\ (solver)\end{tabular} &
	\begin{tabular}[c]{@{}l@{}}AS last time\\ (solver)\end{tabular} \\
		\midrule
		clamp & 683 & 177 & 110 & 12.903 & 99.272 (12.099) & 17.110 (0.941)
		& 1.880 (0.282) & 82.163 (11.158) & 32.490 (4.253) \\ \midrule
		prev\_pow\_2(\#1) & 32 & 0 & 6 & 4.688 & 6.125 (0.266) & 4.312
		(0.144) & 0.875 (0.053) & 1.812 (0.122) & 0.938 (0.063) \\ \midrule
		abs\_diff(\#2) & 575 & 5 & 75 & 10.517 & 19.981 (1.331) & 12.944
		(0.487) & 1.120 (0.095) & 7.037 (0.844) & 1.843 (0.276) \\ \midrule
		bswap32(\#1) & 115 & 8 & 19 & 8.67 & 16.565 (1.235) & 12.313 (0.984)
		& 1.000 (0.227) & 4.252 (0.251) & 1.226 (0.089) \\ \midrule
		integer\_cmp(\#2) & 93 & 5 & 15 & 9.645 & 21.419 (2.246) & 8.839
		(0.598) & 1.280 (0.275) & 12.581 (1.648) & 4.742 (0.630) \\ \midrule
		even(\#1) & 3 & 2 & 3 & 5.667 & 11.333 (0.558) & 7.000 (0.312) &
		2.333 (0.218) & 4.333 (0.246) & 2.333 (0.154) \\ \midrule
		div255(\#1) & 4 & 0 & 2 & 5 & 6.500 (0.262) & 4.000 (0.143) & 0.750
		(0.051) & 2.500 (0.119) & 1.500 (0.068) \\ \midrule
		reverse\_bits(\#1) & 276 & 0 & 11 & 8.978 & 25.264 (2.926) & 16.192
		(0.678) & 1.978 (0.112) & 9.072 (2.248) & 1.895 (0.454) \\ \midrule
		binary\_log(\#1) & 48 & 0 & 5 & 6.708 & 23.562 (5.870) & 10.938
		(2.191) & 2.125 (0.728) & 12.625 (3.679) & 8.750 (3.235) \\ \midrule
		median(\#3) & 332 & 42 & 60 & 13.669 & 119.226 (26.739) & 17.789
		(1.323) & 2.250 (0.454) & 101.437 (25.416) & 33.931 (8.548) \\
		\midrule
		hex\_value(\#1) & 0 & 0 & 0 & 0 & 0.000 (0.000) & 0.000 (0.000) &
		0.000 (0.000) & 0.000 (0.000) & 0.000 (0.000) \\ \midrule
		\begin{tabular}[c]{@{}l@{}}get\_descriptor\_\\
		length\_24b(\#1)\end{tabular} & 22 & 9 & 2 & 9 & 16.682 (0.583) &
			11.909 (0.328) & 1.136 (0.091) & 4.773 (0.255) & 1.591 (0.098) \\
			\midrule
			tile\_pos(\#4) & 5617 & 407 & 909 & 10.902 & 53.478 (23.124) &
			10.968 (1.767) & 2.836 (1.409) & 42.510 (21.357) & 18.090 (10.019)
			\\ \midrule
			\begin{tabular}[c]{@{}l@{}}dirac\_picture\_n\_\\
			before\_m(\#2)\end{tabular} & 330 & 2 & 18 & 13.224 & 25.736
				(2.974) & 13.048 (0.638) & 0.855 (0.084) & 12.688 (2.335) &
				2.124 (0.386) \\ \midrule
				ps\_id\_to\_tk(\#1) & 0 & 0 & 0 & 0 & 0.000 (0.000) & 0.000
				(0.000) & 0.000 (0.000) & 0.000 (0.000) & 0.000 (0.000) \\
				\midrule
				leading\_zero\_count(\#1) & 41 & 0 & 7 & 18.561 & 39.000 (4.529)
				& 22.780 (1.174) & 1.000 (0.146) & 16.220 (3.355) & 2.488
				(0.721) \\ \midrule
				trailing\_zero\_count(\#1) & 46 & 0 & 4 & 5.87 & 16.196 (3.832)
				& 9.109 (1.097) & 2.065 (0.738) & 7.087 (2.735) & 3.478 (1.322)
				\\ \midrule
				popcnt\_32(\#1) & 0 & 0 & 0 & 0 & 0.000 (0.000) & 0.000 (0.000)
				& 0.000 (0.000) & 0.000 (0.000) & 0.000 (0.000) \\ \midrule
				parity(\#1) & 0 & 0 & 0 & 0 & 0.000 (0.000) & 0.000 (0.000) &
				0.000 (0.000) & 0.000 (0.000) & 0.000 (0.000) \\ \midrule
				dv\_audio\_12\_to\_16(\#1) & 0 & 0 & 0 & 0 & 0.000 (0.000) &
				0.000 (0.000) & 0.000 (0.000) & 0.000 (0.000) & 0.000 (0.000) \\
				\midrule
				is\_power\_2(\#1) & 0 & 0 & 0 & 0 & 0.000 (0.000) & 0.000
				(0.000) & 0.000 (0.000) & 0.000 (0.000) & 0.000 (0.000) \\
				\midrule
				RenderRGB(\#3) & 763 & 2 & 64 & 10.814 & 27.469 (1.518) & 17.021
				(0.814) & 1.046 (0.143) & 10.448 (0.704) & 2.819 (0.221) \\
				\midrule
				decode\_BCD(\#1) & 0 & 0 & 0 & 0 & 0.000 (0.000) & 0.000 (0.000)
				& 0.000 (0.000) & 0.000 (0.000) & 0.000 (0.000) \\ \midrule
				\begin{tabular}[c]{@{}l@{}}mpga\_get\_\\
				frame\_samples(\#1)\end{tabular} & 22 & 15 & 4 & 5 & 7.909
					(0.887) & 5.273 (0.505) & 1.182 (0.361) & 2.636 (0.381) &
					1.773 (0.345) \\ \bottomrule
				\end{tabular}
			\end{table*}

%% file: inequiv_full_table.tex
\begin{table*}[]
\centering
\caption{Metrics for the insubstitutable conclusion for all reference functions}
\label{table:inequiv_full}
	\begin{tabular}{@{}llllllll@{}}
		\toprule
		\multicolumn{1}{l}{fn\_name} & \multicolumn{1}{l}{\#} &
		\multicolumn{1}{l}{steps} &
		\multicolumn{1}{l}{\begin{tabular}[c]{@{}l@{}}total time\\
		(solver)\end{tabular}} &
				\multicolumn{1}{l}{\begin{tabular}[c]{@{}l@{}}CE total time \\
				(solver)\end{tabular}} &
						\multicolumn{1}{l}{\begin{tabular}[c]{@{}l@{}}CE last time
						\\ (solver)\end{tabular}} &
								\multicolumn{1}{l}{\begin{tabular}[c]{@{}l@{}}AS total
								time \\ (solver)\end{tabular}} &
										\multicolumn{1}{l}{\begin{tabular}[c]{@{}l@{}}AS
										last time\\ (solver)\end{tabular}} \\ \midrule
										\multicolumn{1}{l}{clamp} &
										\multicolumn{1}{l}{40553} &
										\multicolumn{1}{l}{7.711} &
										\multicolumn{1}{l}{63.015 (6.361)} &
										\multicolumn{1}{l}{8.171 (0.375)} &
										\multicolumn{1}{l}{1.703 (0.112)} &
										\multicolumn{1}{l}{54.844 (5.986)} &
										\multicolumn{1}{l}{38.464 (4.032)} \\ \midrule
										\multicolumn{1}{l}{prev\_pow\_2(\#1)} &
										\multicolumn{1}{l}{46767} &
										\multicolumn{1}{l}{4.258} &
										\multicolumn{1}{l}{7.521 (0.492)} &
										\multicolumn{1}{l}{4.833 (0.225)} &
										\multicolumn{1}{l}{2.008 (0.154)} &
										\multicolumn{1}{l}{2.687 (0.267)} &
										\multicolumn{1}{l}{1.502 (0.201)} \\ \midrule
										\multicolumn{1}{l}{abs\_diff(\#2)} &
										\multicolumn{1}{l}{46250} &
										\multicolumn{1}{l}{8.205} &
										\multicolumn{1}{l}{18.735 (1.384)} &
										\multicolumn{1}{l}{11.281 (0.411)} &
										\multicolumn{1}{l}{2.281 (0.124)} &
										\multicolumn{1}{l}{7.453 (0.973)} &
										\multicolumn{1}{l}{3.268 (0.562)} \\ \midrule
										\multicolumn{1}{l}{bswap32(\#1)} &
										\multicolumn{1}{l}{46708} &
										\multicolumn{1}{l}{4.682} &
										\multicolumn{1}{l}{8.184 (0.493)} &
										\multicolumn{1}{l}{5.136 (0.196)} &
										\multicolumn{1}{l}{1.764 (0.102)} &
										\multicolumn{1}{l}{3.048 (0.297)} &
										\multicolumn{1}{l}{1.620 (0.217)} \\ \midrule
										\multicolumn{1}{l}{integer\_cmp(\#2)} &
										\multicolumn{1}{l}{46467} &
										\multicolumn{1}{l}{5.249} &
										\multicolumn{1}{l}{15.324 (1.772)} &
										\multicolumn{1}{l}{7.850 (0.404)} &
										\multicolumn{1}{l}{2.816 (0.177)} &
										\multicolumn{1}{l}{7.474 (1.369)} &
										\multicolumn{1}{l}{4.640 (0.999)} \\ \midrule
										\multicolumn{1}{l}{even(\#1)} &
										\multicolumn{1}{l}{46823} &
										\multicolumn{1}{l}{4.218} &
										\multicolumn{1}{l}{12.699 (0.859)} &
										\multicolumn{1}{l}{7.088 (0.229)} &
										\multicolumn{1}{l}{2.883 (0.149)} &
										\multicolumn{1}{l}{5.611 (0.630)} &
										\multicolumn{1}{l}{3.881 (0.529)} \\ \midrule
										\multicolumn{1}{l}{div255(\#1)} &
										\multicolumn{1}{l}{46823} &
										\multicolumn{1}{l}{4.381} &
										\multicolumn{1}{l}{7.568 (0.463)} &
										\multicolumn{1}{l}{4.849 (0.206)} &
										\multicolumn{1}{l}{1.824 (0.117)} &
										\multicolumn{1}{l}{2.719 (0.257)} &
										\multicolumn{1}{l}{1.499 (0.196)} \\ \midrule
										\multicolumn{1}{l}{reverse\_bits(\#1)} &
										\multicolumn{1}{l}{46541} &
										\multicolumn{1}{l}{12.536} &
										\multicolumn{1}{l}{50.866 (5.645)} &
										\multicolumn{1}{l}{22.051 (0.784)} &
										\multicolumn{1}{l}{2.359 (0.103)} &
										\multicolumn{1}{l}{28.815 (4.861)} &
										\multicolumn{1}{l}{12.573 (1.454)} \\ \midrule
										\multicolumn{1}{l}{binary\_log(\#1)} &
										\multicolumn{1}{l}{46528} &
										\multicolumn{1}{l}{4.024} &
										\multicolumn{1}{l}{25.631 (6.368)} &
										\multicolumn{1}{l}{4.848 (0.551)} &
										\multicolumn{1}{l}{2.004 (0.136)} &
										\multicolumn{1}{l}{20.783 (5.817)} &
										\multicolumn{1}{l}{15.253 (4.314)} \\ \midrule
										\multicolumn{1}{l}{median(\#3)} &
										\multicolumn{1}{l}{32171} &
										\multicolumn{1}{l}{6.484} &
										\multicolumn{1}{l}{89.779 (15.126)} &
										\multicolumn{1}{l}{6.598 (0.312)} &
										\multicolumn{1}{l}{1.723 (0.097)} &
										\multicolumn{1}{l}{83.181 (14.815)} &
										\multicolumn{1}{l}{75.092 (13.180)} \\ \midrule
										\multicolumn{1}{l}{hex\_value(\#1)} &
										\multicolumn{1}{l}{46354} &
										\multicolumn{1}{l}{3.157} &
										\multicolumn{1}{l}{9.233 (2.092)} &
										\multicolumn{1}{l}{4.412 (0.370)} &
										\multicolumn{1}{l}{2.333 (0.128)} &
										\multicolumn{1}{l}{4.821 (1.722)} &
										\multicolumn{1}{l}{3.894 (1.471)} \\ \midrule
										\multicolumn{1}{l}{transform\_from\_basic\_ops(\#10)}
										& \multicolumn{1}{l}{40169} &
										\multicolumn{1}{l}{10.253} &
										\multicolumn{1}{l}{115.732 (8.667)} &
										\multicolumn{1}{l}{9.020 (0.452)} &
										\multicolumn{1}{l}{1.552 (0.079)} &
										\multicolumn{1}{l}{106.712 (8.215)} &
										\multicolumn{1}{l}{75.875 (5.514)} \\ \midrule
										\multicolumn{1}{l}{get\_descriptor\_length\_24b(\#1)}
										& \multicolumn{1}{l}{46625} &
										\multicolumn{1}{l}{5.442} &
										\multicolumn{1}{l}{11.687 (0.718)} &
										\multicolumn{1}{l}{7.791 (0.329)} &
										\multicolumn{1}{l}{2.384 (0.104)} &
										\multicolumn{1}{l}{3.896 (0.388)} &
										\multicolumn{1}{l}{1.988 (0.301)} \\ \midrule
										\multicolumn{1}{l}{tile\_pos(\#4)} &
										\multicolumn{1}{l}{24696} &
										\multicolumn{1}{l}{8.031} &
										\multicolumn{1}{l}{67.636 (27.126)} &
										\multicolumn{1}{l}{7.045 (0.397)} &
										\multicolumn{1}{l}{1.756 (0.091)} &
										\multicolumn{1}{l}{60.591 (26.728)} &
										\multicolumn{1}{l}{46.309 (20.400)} \\ \midrule
										\multicolumn{1}{l}{diract\_picture\_n\_before\_m(\#2)}
										& \multicolumn{1}{l}{46393} &
										\multicolumn{1}{l}{6.615} &
										\multicolumn{1}{l}{15.315 (1.327)} &
										\multicolumn{1}{l}{6.968 (0.315)} &
										\multicolumn{1}{l}{2.226 (0.116)} &
										\multicolumn{1}{l}{8.347 (1.012)} &
										\multicolumn{1}{l}{3.746 (0.337)} \\ \midrule
										\multicolumn{1}{l}{ps\_id\_to\_tk(\#1)} &
										\multicolumn{1}{l}{46721} &
										\multicolumn{1}{l}{4.41} &
										\multicolumn{1}{l}{15.811 (2.370)} &
										\multicolumn{1}{l}{7.414 (1.090)} &
										\multicolumn{1}{l}{2.579 (0.190)} &
										\multicolumn{1}{l}{8.397 (1.280)} &
										\multicolumn{1}{l}{6.504 (1.127)} \\ \midrule
										\multicolumn{1}{l}{leading\_zero\_count(\#1)} &
										\multicolumn{1}{l}{46727} &
										\multicolumn{1}{l}{7.838} &
										\multicolumn{1}{l}{16.737 (2.105)} &
										\multicolumn{1}{l}{8.462 (0.598)} &
										\multicolumn{1}{l}{2.090 (0.136)} &
										\multicolumn{1}{l}{8.275 (1.507)} &
										\multicolumn{1}{l}{3.473 (0.609)} \\ \midrule
										\multicolumn{1}{l}{trailing\_zero\_count(\#1)} &
										\multicolumn{1}{l}{46701} &
										\multicolumn{1}{l}{3.392} &
										\multicolumn{1}{l}{19.508 (6.189)} &
										\multicolumn{1}{l}{4.161 (0.706)} &
										\multicolumn{1}{l}{1.881 (0.135)} &
										\multicolumn{1}{l}{15.347 (5.483)} &
										\multicolumn{1}{l}{13.786 (5.088)} \\ \midrule
										\multicolumn{1}{l}{popcnt\_32(\#1)} &
										\multicolumn{1}{l}{46802} &
										\multicolumn{1}{l}{5.602} &
										\multicolumn{1}{l}{11.500 (0.818)} &
										\multicolumn{1}{l}{7.296 (0.313)} &
										\multicolumn{1}{l}{2.471 (0.155)} &
										\multicolumn{1}{l}{4.204 (0.504)} &
										\multicolumn{1}{l}{2.076 (0.335)} \\ \midrule
										\multicolumn{1}{l}{parity(\#1)} &
										\multicolumn{1}{l}{46821} &
										\multicolumn{1}{l}{4.988} &
										\multicolumn{1}{l}{9.968 (0.644)} &
										\multicolumn{1}{l}{6.447 (0.292)} &
										\multicolumn{1}{l}{2.584 (0.179)} &
										\multicolumn{1}{l}{3.521 (0.352)} &
										\multicolumn{1}{l}{1.813 (0.244)} \\ \midrule
										\multicolumn{1}{l}{dv\_audio\_12\_to\_16(\#1)} &
										\multicolumn{1}{l}{46637} &
										\multicolumn{1}{l}{3.884} &
										\multicolumn{1}{l}{17.708 (2.780)} &
										\multicolumn{1}{l}{8.279 (0.598)} &
										\multicolumn{1}{l}{3.607 (0.155)} &
										\multicolumn{1}{l}{9.429 (2.182)} &
										\multicolumn{1}{l}{7.004 (1.673)} \\ \midrule
										\multicolumn{1}{l}{is\_power\_2(\#1)} &
										\multicolumn{1}{l}{46801} &
										\multicolumn{1}{l}{3.791} &
										\multicolumn{1}{l}{9.130 (1.357)} &
										\multicolumn{1}{l}{5.420 (0.316)} &
										\multicolumn{1}{l}{2.819 (0.225)} &
										\multicolumn{1}{l}{3.710 (1.042)} &
										\multicolumn{1}{l}{2.218 (0.659)} \\ \midrule
										\multicolumn{1}{l}{RenderRGB(\#3)} &
										\multicolumn{1}{l}{46061} &
										\multicolumn{1}{l}{5.663} &
										\multicolumn{1}{l}{17.038 (0.901)} &
										\multicolumn{1}{l}{9.718 (0.366)} &
										\multicolumn{1}{l}{2.670 (0.172)} &
										\multicolumn{1}{l}{7.320 (0.535)} &
										\multicolumn{1}{l}{4.023 (0.330)} \\ \midrule
										\multicolumn{1}{l}{decode\_BCD(\#1)} &
										\multicolumn{1}{l}{46824} &
										\multicolumn{1}{l}{4.706} &
										\multicolumn{1}{l}{8.751 (1.124)} &
										\multicolumn{1}{l}{5.516 (0.356)} &
										\multicolumn{1}{l}{1.890 (0.202)} &
										\multicolumn{1}{l}{3.235 (0.768)} &
										\multicolumn{1}{l}{1.903 (0.618)} \\ \midrule
										mpga\_get\_frame\_samples(\#1) & 46235 & 3.366 &
										9.288 (2.057) & 4.887 (0.497) & 2.580 (0.148) &
										4.401 (1.560) & 3.595 (1.454) \\ \bottomrule
									\end{tabular}
								\end{table*}

%% file: timeouts_full_table.tex
\begin{table*}[]
\centering
\caption{Metrics for the timeout conclusion for all reference functions}
\label{table:timeouts_full}
	\begin{tabular}{@{}lllllllll@{}}
		\toprule
		fn\_name & \# & steps & \begin{tabular}[c]{@{}l@{}}total time \\
	(solver)\end{tabular} & \begin{tabular}[c]{@{}l@{}}CE total time \\
	(solver)\end{tabular} & \begin{tabular}[c]{@{}l@{}}CE last time \\
	(solver)\end{tabular} & \begin{tabular}[c]{@{}l@{}}AS total time \\
	(solver)\end{tabular} & \begin{tabular}[c]{@{}l@{}}AS last time \\
	(solver)\end{tabular} & \begin{tabular}[c]{@{}l@{}}AS-stops/\\
	CE-stops\end{tabular} \\ \midrule
	clamp & 5595 & 16.505 & 300.000 (44.278) & 27.856 (8.112) & 9.392
	(6.966) & 272.144 (36.167) & 140.702 (17.457) & 5416/179 \\ \midrule
	prev\_pow\_2(\#1) & 32 & 1 & 300.000 (289.445) & 300.000 (289.445) &
	300.000 (289.445) & 0.000 (0.000) & 0.000 (0.000) & 0/32 \\ \midrule
	abs\_diff(\#2) & 6 & 5.667 & 300.000 (286.525) & 297.333 (286.318) &
	288.167 (285.378) & 2.667 (0.206) & 1.167 (0.112) & 0/6 \\ \midrule
	bswap32(\#1) & 8 & 2.75 & 300.000 (293.526) & 299.125 (293.479) &
	296.250 (293.329) & 0.875 (0.047) & 0.875 (0.047) & 0/8 \\ \midrule
	integer\_cmp(\#2) & 271 & 3.085 & 300.000 (247.247) & 296.347
	(246.627) & 288.122 (243.312) & 3.653 (0.620) & 1.063 (0.209) & 3/268
	\\ \midrule
	even(\#1) & 5 & 1.8 & 300.000 (116.452) & 299.600 (116.434) & 297.400
	(116.320) & 0.400 (0.019) & 0.400 (0.019) & 0/5 \\ \midrule
	div255(\#1) & 4 & 2.5 & 300.000 (294.241) & 299.500 (294.203) &
	297.500 (294.115) & 0.500 (0.037) & 0.500 (0.037) & 0/4 \\ \midrule
	reverse\_bits(\#1) & 14 & 3 & 300.000 (292.294) & 298.714 (292.182) &
	294.786 (291.965) & 1.286 (0.112) & 1.286 (0.112) & 0/14 \\ \midrule
	binary\_log(\#1) & 255 & 1.239 & 300.000 (207.291) & 298.824 (206.920)
	& 277.769 (203.879) & 1.176 (0.371) & 0.949 (0.336) & 19/236 \\
	\midrule
	median(\#3) & 14328 & 13.634 & 300.000 (65.444) & 15.655 (2.144) &
	3.266 (1.319) & 284.345 (63.300) & 167.910 (35.663) & 14184/144 \\
	\midrule
	hex\_value(\#1) & 477 & 1.013 & 300.000 (268.765) & 299.964 (268.754)
	& 298.753 (268.165) & 0.036 (0.010) & 0.027 (0.007) & 2/475 \\
	\midrule
	\begin{tabular}[c]{@{}l@{}}transform\_from\_\\
	basic\_ops(\#10)\end{tabular} & 6409 & 18.381 & 300.000 (27.949) &
		22.098 (3.092) & 4.510 (2.408) & 277.902 (24.857) & 172.895 (14.278)
		& 6319/90 \\ \midrule
		\begin{tabular}[c]{@{}l@{}}get\_descriptor\_\\
		length\_24b(\#1)\end{tabular} & 184 & 1.391 & 300.000 (233.380) &
			299.832 (233.373) & 298.853 (233.277) & 0.168 (0.006) & 0.168
			(0.006) & 0/184 \\ \midrule
			tile\_pos(\#4) & 16518 & 7.634 & 300.000 (280.532) & 8.118 (1.326)
			& 2.782 (0.988) & 291.882 (279.206) & 256.574 (249.372) & 16441/77
			\\ \midrule
			\begin{tabular}[c]{@{}l@{}}dirac\_picture\_\\
			n\_before\_m(\#2)\end{tabular} & 108 & 51.481 & 300.000 (137.988)
				& 132.556 (87.679) & 89.917 (85.144) & 167.444 (50.309) & 25.954
				(3.204) & 74/34 \\ \midrule
				ps\_id\_to\_tk(\#1) & 110 & 1.118 & 300.000 (258.764) & 299.755
				(258.748) & 291.764 (250.903) & 0.245 (0.015) & 0.218 (0.014) &
				3/107 \\ \midrule
				\begin{tabular}[c]{@{}l@{}}leading\_zero\_\\
				count(\#1)\end{tabular} & 63 & 5.079 & 300.000 (143.608) &
					297.254 (143.230) & 171.063 (111.259) & 2.746 (0.379) & 0.841
					(0.100) & 1/62 \\ \midrule
					\begin{tabular}[c]{@{}l@{}}trailing\_zero\_\\
					count(\#1)\end{tabular} & 84 & 1.476 & 300.000 (283.679) &
						299.155 (283.545) & 285.417 (270.053) & 0.845 (0.134) &
						0.643 (0.111) & 4/80 \\ \midrule
						popcnt\_32(\#1) & 29 & 1 & 300.000 (295.366) & 300.000
						(295.366) & 300.000 (295.366) & 0.000 (0.000) & 0.000
						(0.000) & 0/29 \\ \midrule
						parity(\#1) & 10 & 1.2 & 300.000 (266.296) & 299.900
						(266.293) & 275.100 (266.280) & 0.100 (0.003) & 0.100
						(0.003) & 0/10 \\ \midrule
						\begin{tabular}[c]{@{}l@{}}dv\_audio\_\\
						12\_to\_16(\#1)\end{tabular} & 194 & 1.026 & 300.000
							(290.336) & 299.979 (290.334) & 296.928 (288.827) & 0.021
							(0.002) & 0.021 (0.002) & 1/193 \\ \midrule
							is\_power\_2(\#1) & 30 & 3.667 & 300.000 (291.082) &
							297.833 (290.309) & 293.867 (290.012) & 2.167 (0.773) &
							1.133 (0.375) & 0/30 \\ \midrule
							RenderRGB(\#3) & 7 & 4.429 & 300.000 (115.721) & 297.000
							(115.538) & 290.714 (115.275) & 3.000 (0.184) & 1.714
							(0.099) & 0/7 \\ \midrule
							decode\_BCD(\#1) & 7 & 1.857 & 300.000 (126.084) & 299.714
							(126.040) & 298.429 (125.986) & 0.286 (0.044) & 0.286
							(0.044) & 0/7 \\ \midrule
							\begin{tabular}[c]{@{}l@{}}mpga\_get\_\\
							frame\_samples(\#1)\end{tabular} & 574 & 1.024 & 300.000
								(289.464) & 299.963 (289.460) & 297.423 (288.201) &
								0.037 (0.003) & 0.035 (0.003) & 4/570 \\ \bottomrule
							\end{tabular}
						\end{table*}